\documentclass[aps,prb,amsmath,amssymb,reprint,superscriptaddress,preprintnumbers,showpacs,intlimits]{revtex4-1}

\usepackage{bm,mathrsfs}
\usepackage[mathcal]{euscript}
\usepackage[breaklinks=true,unicode=true,urlcolor = blue,colorlinks = true,citecolor = blue,linkcolor = blue]{hyperref}
\usepackage{float}
\usepackage{graphicx}
\usepackage{subcaption}
\usepackage[table]{xcolor}
\usepackage[normalem]{ulem}

\renewcommand{\vec}[1]{\bm{#1}}
\usepackage{tikz}
\usepackage{tikz-3dplot}
\usetikzlibrary{arrows,shapes,backgrounds}
\usetikzlibrary{positioning,arrows}
\usetikzlibrary{calc,fadings,decorations.pathreplacing}
\tikzset{%
  >=latex, 
  inner sep=0pt,%
  outer sep=2pt,%
  mark coordinate/.style={inner sep=0pt,outer sep=0pt,minimum size=3pt,
    fill=black,circle}%
}

\begin{document}
\title{Vortex Polarity Switching in Magnets with Surface Anisotropy}

\author{Oleksandr V. Pylypovskyi}
\email[Corresponding author. Electronic address:]{engraver@univ.net.ua}
\affiliation{Taras Shevchenko National University of Kyiv, 01601 Kyiv, Ukraine}

\author{Denis D. Sheka}
\email{sheka@univ.net.ua}
\affiliation{Taras Shevchenko National University of Kyiv, 01601 Kyiv, Ukraine}

\author{Volodymyr P. Kravchuk}
 \email{vkravchuk@bitp.kiev.ua}
 \affiliation{Institute for Theoretical Physics, 03143 Kyiv, Ukraine}

\author{Yuri Gaididei}
\email{ybg@bitp.kiev.ua}
\affiliation{Institute for Theoretical Physics, 03143 Kyiv, Ukraine}

\date{January 20, 2015}

%
%

\begin{abstract}
Vortex core reversal in magnetic particle is essentially influenced by a surface anisotropy. Under the action of a perpendicular static magnetic field the vortex core undergoes a shape deformationof pillow- or barrel-shaped type, depending on the type of the surface anisotropy. This deformation plays a key point in the switching mechanism: We predict that the vortex polarity switching is accompanied (i) by a linear singularity in case of Heisenberg magnet with bulk anisotropy only and (ii) by a point singularities in case of surface anisotropy or exchange anisotropy. We study in details the switching process using spin-lattice simulations and propose a simple analytical description using a \emph{wired core model}, which provides an adequate description of the Bloch point statics, its \emph{dynamics} and the Bloch point mediated switching process. Our analytical predictions are
confirmed by spin-lattice simulations for Heisenberg magnet and micromagnetic simulations for nanomagnet with account of a dipolar interaction.
\end{abstract}




\maketitle

\section{Introduction}

Magnetization reversal in small magnetic particles is one of the fundamental issues of the modern magnetism. Different concepts of switching, including switching by magnetic fields (quasistatic and precessional ones), switching by spin injection and all optical switching are intensively studied and widely used for applications in magnetic data storage.\cite{Stohr06,Hillebrands06} Nowadays inhomogeneous switching,\cite{Kronmueller07v2} i.~e. the process of magnetization reversal for inhomogeneous magnetization configurations, attracts growing interest. Particular attention is paid to topologically protected configuration like magnetic vortex. The vortex is determined by a planar closed flux-free configuration in a sample with a localized nonplanar magnetization, so called vortex core. The sense of the core magnetization direction is characterized by the vortex polarity (up, $p=+1$ or down, $p=-1$). Because of the topological stability of vortices with different polarities, one can consider the vortex
polarity as a bit of information in nonvolatile magnetic vortex random-access memories. \cite{Kim08,Pigeau10,Yu11a} The corresponding inhomogeneous magnetization reversal in the vortex-state particles is called the vortex core reversal or the vortex polarity switching.

There exist several concepts of the vortex polarity switching, e.~g. by magnetic fields, by spin injection, see Ref.~\onlinecite{Gaididei08b} and references therein. Independently of these concepts one can separate two main scenarios: axially-symmetric (or punch-through) scenario and axially-asymmetric one.  In the first scenario the vortex does not make the macroscopic motion, the vortex core reversal is caused by the direct excitation of radially symmetric magnon modes. Such a switching occurs, e.~g. under the action of dc transversal magnetic field \cite{Okuno02,Hoellinger03, Thiaville03, Kravchuk07a}, or by ac transversal fields \cite{Wang12,Yoo12,Pylypovskyi13b,Pylypovskyi13e,Helsen14}. The axially-asymmetric scenario is caused by nonlinear resonance between coupled magnon modes;\cite{Kravchuk09, Gaididei10b} it is accompanied by the temporary creation of the vortex-antivortex pairs\cite{Waeyenberge06,Hertel07}. Such a picture was observed in different setups.\cite{Gaididei08b}

Here we consider the axially-symmetrical switching of the vortex polarity. During the reversal process the main changes occur in a circular Bloch line\cite{Hubert98}, which is the line passing trough the sample thickness and connecting centers of all vortex cross-sections. The magnetization direction in the Bloch line corresponds to the vortex polarity: it has to reverse its direction during the switching process. In case when a reversal occurs simultaneously for the whole Bloch line $\vec{\gamma }$, there appears a line singularity during switching: the magnetization vector $\vec{m}$ vanishes at the switching moment for the whole line, $\vec{m}(\vec{\gamma })=0$, and the temporary pure planar vortex (\emph{linear singularity}) appears during the switching. Such a two-dimensional (2D) switching process was considered in Ref.~\onlinecite{Kravchuk07a} to describe the reversal in very thin magnets. Another possibility was suggested by \citet{Arrott79}.
Arrott's model is stray-field-free, where the switching is made possible by propagating of point singularities, Bloch points with coordinates $\vec{r}_{\text{BP}}$.\cite{Arrott79,Hubert98a,Hubert98} In that case the reversal occurs nonsimultaneously: the magnetization vanishes only in one or a few points, the so-called Bloch points, $\vec{m}(\vec{r}_{\text{BP}})=0$, and the switching is accompanied by the nucleation, motion and final annihilation of the Bloch points. Such a process was studied by \citet{Thiaville03} using micromagnetic simulations: the switching process is typically accompanied by the creation of two Bloch points, however, the single Bloch point scenario was also mentioned.\cite{Thiaville03}

The purpose of the current study is to describe process of the axially-symmetrical switching, to find the mechanism, which is responsible for simultaneous or nonsimultaneous switching, to understand how many Bloch points are nucleated during the switching, and to describe this process analytically.

Point singularities were introduced in magnetism by \citet{Feldtkeller65b} and \citet{Doering68}. For example, in the vicinity of the Bloch point, situated at $\vec{r}_{\text{BP}}=0$, the magnetization distribution has a hedgehog-like configuration, $\vec m = \pm Q\vec{r}/r$ with $Q$ being the constant orthogonal matrix.\cite{Borisov11a} They are analogues of magnetic monopoles in elementary particle physics for spin waves.\cite{Elias14} Bloch points were observed as twisting of the Bloch lines in garnet crystals \cite{Kabanov88,Kabanov89} and as singular point of the Bloch point domain wall in magnetic nanowires \cite{DaCol14}. According to simulations the Bloch points appear as transient states in the vortex dynamics under the action of dc \cite{Thiaville03} and ac \cite{Yoo12,Pylypovskyi13e} transversal fields; as temporal formations (micromagnetic drops) during the magnetization reversal in soft magnetic cylinders.\cite{Hertel04a} The Bloch points 
are also created during the vortex-antivortex annihilation
processes. \cite{Hertel06}

Theoretical treatment of the Bloch point is a complicated task. Even a ``simple'' problem of the Bloch point structure in a spherical particle caused a long-time discussion, see Ref.~\onlinecite{Pylypovskyi12} and references therein. The problem becomes complicated due to the fact that the Bloch point distribution with $\vec{m}(\vec{r}_{\text{BP}})=0$ does not fulfil the condition of magnetization normalization, $|\vec{m}|=1$. There is no physical singularity in reality. Due to the discreetness of the spin lattice, the Bloch point (as well as linear singularity) is situated in an interstitial site position, hence the singularity appears only in continuum description. That is why the modeling of Bloch points is also nontrivial task:
there appear mesh dependent problems, e.g. mesh-friction effect and a strong mesh dependence of the switching field.\cite{Thiaville03} Therefore specific simulators with atomistic resolution are needed to overcome such difficulties.\cite{Pylypovskyi12,Andreas14} The role of discreetness of the magnetic lattice becomes of great importance; in particular, namely the lattice creates an effective pinning field for the Bloch point;\cite{Kim13a} it changes the Bloch point behavior relatively to the main crystallographic directions when a depinning field for one direction can be much lower than in  other direction.\cite{Andreas14}

The problem of dynamics of Bloch points is not less challenging. The spectrum of Bloch point oscillations along the Bloch line was measured for yttrium garnet ferrite.\cite{Gornakov89} Dynamical properties of the Bloch point, its mobility and mass were derived for the Bloch point inside the Bloch line.\cite{Malozemoff79,Kufaev88,Kufaev89,Galkina93} We are interesting in a specific dynamics the Bloch point, namely, the Bloch point mediated switching of the vortex polarity. It is already known from our recent studies \cite{Pylypovskyi13b,Pylypovskyi13e} that the dominating contribution to the switching mechanism is caused by the exchange interaction inside the vortex core. Therefore one can expect that the dipolar interaction, which is essential for the statics and slow dynamics of the magnet, in particular, for the vortex state nanodot, does not play the dominant role for the switching process.

Argued by above mentioned reasons of the dominant role of exchange interaction we do not take into account in most cases the dipolar interaction. Instead we additionally include into consideration the surface anisotropy effects. In a disk-shaped particles the surface anisotropy favors magnetization curling states of onion (two half vortex \cite{Kireev03} or capacitor \cite{Leonel07}) configuration and a vortex one \cite{Kireev03,Leonel07}. Below we consider the vortex configuration. One has to note that the vortex is the metastable state in model under consideration,\cite{Kireev03} while it can form the ground state in models with finite skin depth of the surface anisotropy\cite{Leonel07} or in models with dipolar induced effective inhomogeneous anisotropy.\cite{Caputo07b} In the last case there appear two effective inhomogeneous anisotropy terms: one is effective easy-plane anisotropy of face surface charges and another one is effective anisotropy of edge surface charges; \cite{Caputo07b} namely these two
effective anisotropies are responsible for the the vortex state configuration.

In the current study we consider the Heisenberg disk-shaped magnet in a vortex configuration. Without surface anisotropy the vortex profile does not depend on a thickness coordinate $z$, it has a 2D shape. The presence of the surface anisotropy breaks such a symmetry and results in deformation of the vortex core profile. Depending on the type of surface anisotropy, the vortex profile becomes barrel- or pillow-shaped deformed for the easy-surface (ES) and easy-normal (EN) surface anisotropies, respectively.\cite{Pylypovskyi14} We will see below that inhomogeneity of the Bloch line is crucial for understanding the Bloch point mediated switching mechanism: the switching process in Heisenberg three dimensional (3D) magnet with bulk anisotropy only is mediated by the linear singularity. In contrast, the switching in magnet with surface anisotropy is accompanied by the simultaneous nucleation of
two Bloch points on the face surfaces of the sample for the EN case and by the nucleation of two Bloch points inside the sample (at the center of the Bloch line) for the ES case. We also analyze the case of the exchange anisotropy; qualitatively the mechanism is very similar to the EN surface anisotropy: during the switching two Bloch points enter the sample on face surfaces, move inside to each other and finally annihilate at the center of the disk. We study in details the process of the vortex core reversal, compute the Bloch points trajectories, and find how the Bloch point speed depends on the sample thickness. We propose a simple analytical picture which describes both statics of the vortex in the cone phase with account of the surface anisotropy effects and \emph{dynamics} of the vortex polarity switching process, which is in agreement with the full-scale spin-lattice simulations. We also performed micromagnetic OOMMF\cite{oommf} simulations with account of surface anisotropy and additional dipolar
interaction, which confirm main features of our vortex core deformation.

The paper is organized as follows. In the Sec.~\ref{sec:model} we describe the model of the Heisenberg magnet with account of both bulk and surface anisotropies. In the Sec.~\ref{sec:swi} we present results of spin-lattice simulations of vortex polarity switching for Heisenberg magnets with bulk easy-plane single-ion anistropy (Sec.~\ref{sec:homogeneous}), with additional easy-surface and easy-normal surface anisotropies (Sec.~\ref{sec:heisenberg-with-sa}) and with bulk exchange anisotropy (Appendix~\ref{sec:Heisenberg-SAex-numerics}). We make analytical description of the switching phenomenon in the framework of proposed \emph{wired core model} which demostrates a good qualitative agreement with simulations in Sec.~\ref{sec:wcm}. To validate our results for nanoparticles we modeled a vortex state nanodisk using OOMMF micromagnetic simulations with account of dipolar interaction and surface
anisotropy, which are discussed
in Sec.~\ref{sec:discussion}. In Appendix~\ref{app:m0} we consider how the homogeneous state is deformed with account of the surface anisotropy. The basic equations for the wired core model are derived in Appendix~\ref{app:core}.

\section{The Model and the vortex solution}
\label{sec:model}

Let us consider a classical lattice Heisenberg magnet with a simple cubic lattice and Hamiltonian
\begin{subequations} \label{eq:H-total}
\begin{equation} \label{eq:H-total-1}
\mathcal H =  - J\mathcal{S}^2 \sum_{(\vec n, \vec \delta )} \vec m_{\vec n} \cdot \vec m_{\vec n + \vec \delta } - 2\mu_BH\mathcal{S}\sum_{\vec n} (\vec m_{\vec n}\cdot \hat{z})  +   \mathcal{H}^{\text{an}},
\end{equation}
where $J>0$ is the exchange integral, $\mathcal S = 1/2$ is the length of a classical spin, $\vec{m}_{\vec{n}} = (\sqrt{1-m_{\vec{n}}^2}\cos\phi_{\vec{n}}, \sqrt{1-m_{\vec{n}}^2}\sin\phi_{\vec{n}}, m_{\vec{n}})$ is the normalized magnetic moment on a 3D site position $\vec{n}$, the 3D index $\vec \delta$ runs over the nearest neighbors, $\mu_B$ is the Bohr magneton, $H$ is the intensity of external dc magnetic field, directed along $\hat{z}$-axis, and $\mathcal{H}^{\text{an}}$ is the Hamiltonian of single-ion anisotropy. The latter includes the bulk easy-plane anisotropy (with the anisotropy constant $K > 0$), which favors magnetization distribution within $xy$-plane, and the N\'{e}el surface anisotropy term (with the anisotropy constant $K_s$) \cite{Neel54},
\begin{equation} \label{eq:H-EP+SA}
\mathcal{H}^{\text{an}} = \frac{K\mathcal{S}^2}{2}\sum_{\vec n} (\vec m_{\vec n} \cdot \hat{\vec{z}})^2  - \frac{K_s\mathcal{S}^2}{2} \sum_{(\vec{l}, \vec{\delta })} (\vec m_{\vec l} \cdot \vec{u}_{\vec{l}\vec{\delta }})^2.
\end{equation}
\end{subequations}
where index $\vec l$ runs over the surface sites and the unit vector $\vec{u}_{\vec{l}\vec{\delta }}$ connects the nearest neighbors of the lattice. Dynamics of magnetization is governed by the discrete version of the Landau-Lifshitz-Gilbert equation
\begin{equation} \label{eq:LLG}
\frac{\mathrm d\vec m_{\vec n}}{\mathrm d\tau } = \vec m_{\vec n} \times \frac{\partial \mathscr H}{\partial \vec m_{\vec n}} + \varepsilon \vec m_{\vec n} \times \frac{\mathrm d\vec m_{\vec n}}{\mathrm d\tau },
\end{equation}
where $\tau = \Omega t$ is the normalized time, $\Omega = K\mathcal{S}/\hslash$, $\mathscr  H = \mathcal H /(K\mathcal{S}^2)$ is the normalized energy, $\hslash$ is the Plank's constant, and $\varepsilon $ is the Gilbert damping constant.

The continuum approach is based on smoothing of the lattice model using the normalized magnetization
\begin{equation*}
\vec m(\vec r,\tau ) = \left(\sqrt{1-m^2}\cos\phi, \sqrt{1-m^2}\sin\phi, m \right),
\end{equation*}
where $m=m(\vec r,\tau )$ and $\phi = \phi (\vec r,\tau )$. The total energy, which corresponds to the Hamiltonian \eqref{eq:H-total}, takes the form
\begin{equation} \label{eq:Etot}
\begin{split}
E &= \frac{K\mathcal{S}^2}{a^3} \left( \mathscr{E}_v + \mathscr{E}_s \right),\\
\mathscr{E}_v &= \int \mathrm{d}V \left[ -\frac{\ell^2}2 \vec m \cdot \nabla^2\vec m + \frac{(\vec m\cdot \hat{\vec z})^2}2 - \vec m\cdot \vec h\right], \\
\mathscr{E}_s &= \frac{\varkappa a}{2}\int \mathrm{d}S (\vec{m}\cdot \vec{n}_{s})^2.
\end{split}
\end{equation}
Here $\ell = a\sqrt{\lambda }$ is the magnetic length, $a$ is the lattice constant and $\lambda = J/K$, $\varkappa = K_s/K$ is the surface anisotropy measured in units of $K$ and $\vec n_s$ is the normal to the surface. We will consider cases of small surface anisotropy $|\varkappa| < 1$. The parameter $\vec h$ is the field intensity, normalized by the anisotropy field $H_a$,
\begin{equation*} 
\vec h = \vec H/H_a,\qquad H_a=K\mathcal{S}/(2\mu_B).
\end{equation*}
The equilibrium magnetization distribution can be found by variation of the energy functional~\eqref{eq:Etot} which results in the following boundary-value problem:\cite{Brown63,Hubert98}
\begin{subequations} \label{eq:BVP-tot}
\begin{align}\label{eq:BVP-tot-1}
\vec m \times \left[ \ell^2\vec\nabla^2\vec{m} + \vec h - (\vec{m}\cdot \hat{\vec{z}}) \hat{\vec{z}} \right] = 0,\\
 \label{eq:BVP-tot-2}
\ell\frac{\partial \vec{m}}{\partial \vec{n}_s}\Biggr|_{S} = \dfrac{\varkappa }{\sqrt{\lambda }} (\vec{m}\cdot\vec{n}_s)  \left[ (\vec{m}\cdot\vec{n}_s) \vec{m} - \vec{n}_s\right]\Biggr|_{S},
\end{align}
\end{subequations}
Without the surface anisotropy ($\varkappa = 0$) the ground state of the uniaxial magnet is determined by the field intensity. As it follows from \eqref{eq:BVP-tot-1}, the magnetization $z$-component
$m = h=\text{const}$, the boundary condition \eqref{eq:BVP-tot-2} has the Neumann type, $\partial \vec{m}/\partial \vec{n_s}=0$ which is trivially satisfied for the constant $\vec m$ solution.

The presence of the surface anisotropy drastically changes the symmetry of the problem. The boundary conditions \eqref{eq:BVP-tot-2} becomes of the Robin type \cite{Weisstein98}, which is the source of the symmetry breaking of the magnetization structure and causes the effect of magnetization curling. There appears nonhomogeneous ground states with typical examples as hedgehog, throttled, artichoke, onion, vortex configurations, etc.\cite{Kachkachi02,Kireev03,Kachkachi06,Yanes07,Leonel07,Berger08,Noh12}

\begin{figure}
\begin{center}
\begin{tikzpicture}
\node at (0,0) {\includegraphics[width=\columnwidth]{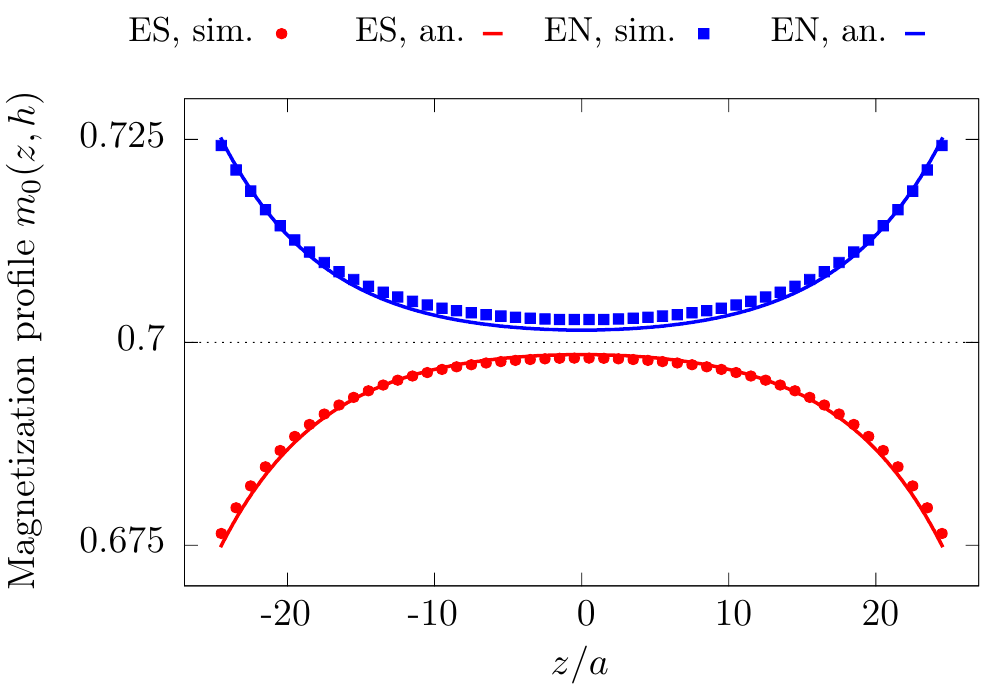}};

\node at (1,-1.2) {\includegraphics[width=0.3\columnwidth]{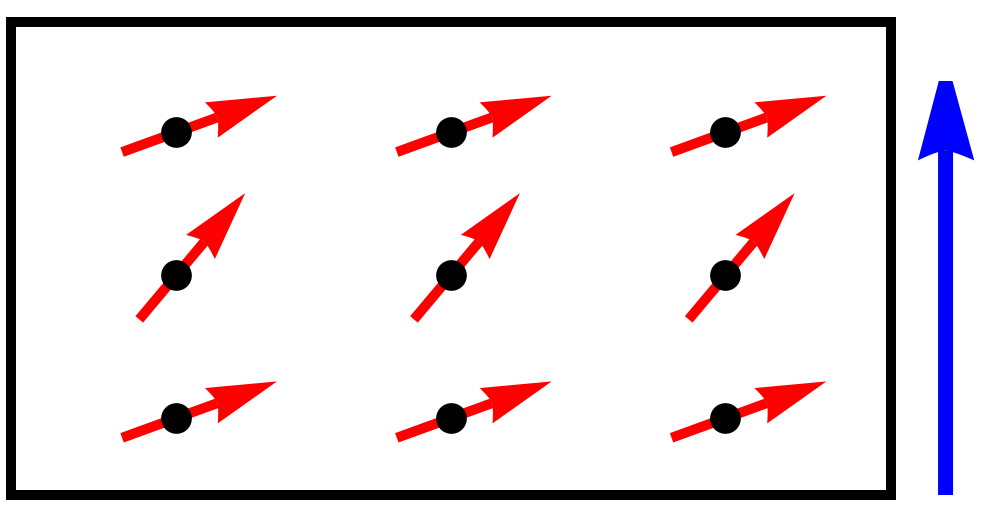}};
\node at (1, 1.2) {\includegraphics[width=0.3\columnwidth]{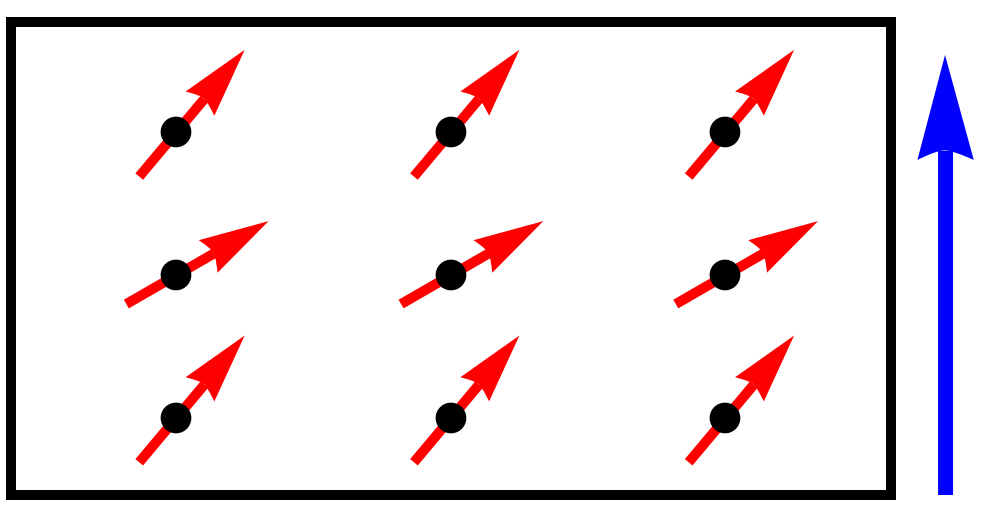}};
\node at (-0.7,-1.2) {(a)};
\node at (-0.7, 1.2) {(b)};
\end{tikzpicture}
\end{center}
\caption{(Color online) Magnetization profile $m_0(z;h)$ as a function the thickness coordinate $z$ for $h=0.7$ (level is shown by dotted line) and $|\kappa| = 0.25$. Symbols correspond to spin-lattice simulations and solid lines to the first order expansion of Eq.~\eqref{eq:m0-z}. Schematics in insets show the magnetization distribution in the film with surface anisotropy for the both cases (red arrows show magnetization direction and blue arrows show field direction).
}
\label{fig:m0}
\end{figure}

Let us start with the easy-plane magnet which has the shape of infinite film of a finite thickness $L=(N-1)a$, where $N$ is the number of lattice sites along thickness coordinate $z$. Without external field and the surface anisotropy the ground state of such a magnet is an uniform easy plane magnetization, $m=0$. The same state takes place with account of the weak surface anisotropy in the no-magnetic field case.

The magnetization configuration under the action of applied field is affected by the surface anisotropy influence: the magnetization distribution breaks its homogeneity in $z$-direction due to the surface anisotropy influence
\begin{equation} \label{eq:m0-z}
m_0(z,h) = h\left[1 - \varkappa \sqrt{\dfrac{1-h^2}{\lambda }} \dfrac{\cosh(\sqrt{1-h^2}z/\ell)}{\sinh(\sqrt{1-h^2}L/2\ell)}\right],
\end{equation}
see Appendix~\ref{app:m0} for details. Typical magnetization profile $m_0$ as function of $z$ for EN and ES surface anisotropies is shown in Fig.~\ref{fig:m0}. Symbols correspond to the spin-lattice simulations,\footnote{The sample of size $100\times 100\times 50$ sites is simulated with magnetic length $\ell=5a$ and $\varkappa=\pm 0.25$ under the action of field $h = 0.7$. The magnetization profile along $\hat{\vec{z}}$ axis is measured in the center of the sample.} see Sec.~\ref{sec:swi} for details.  One can see that the sign of the curvature $m_0=f(z)$ is determined by the sign of the product $\varkappa  h$.

In the following paper we consider the disk-shaped samples with such parameters, that the surface anisotropy favors the vortex configuration. \cite{Kireev03,Leonel07} The in-plane structure of the stationary vortex state $\phi (\vec{r})$ is determined by the following magnetization distribution
\begin{subequations} \label{eq:vortex}
\begin{equation} \label{eq:vortex-phi}
\phi = \chi + C\frac{\pi }{2},
\end{equation}
where $(r ,\chi ,z)$ are the cylinder coordinates. Without surface anisotropy effects, the vortex chirality $C$ can take any value due to the isotropy of Heisenberg exchange, while the surface anisotropy on the edge surface fixes its value: $C=\pm1$ for ES magnets ($\varkappa >0$) and $C=0$ or $C=2$ for EN magnets ($\varkappa <0$). \cite{Pylypovskyi14}

Let us discuss the vortex out-of-plane structure, $m(\vec{r})$. Without magnetic field the out-of-plane vortex profile has a bell-shaped structure\cite{Pylypovskyi14}
\begin{equation} \label{eq:vortex-m4h=0}
m(r ,z)\approx p \exp \left(-\frac{r ^2}{2\ell^2 w^2(z)} \right),
\end{equation}
\end{subequations}
which generalizes a well-known Feldtkeller Ansatz,\cite{Feldtkeller65,Hubert98} originally used for magnets without surface anisotropy, when the dimensionless core width $w\approx1$. Depending on the surface anisotropy type there appears barrel- or pillow- shaped deformation of the vortex core $w(z)$ for $\varkappa > 0$ and $\varkappa < 0$ respectively. \cite{Pylypovskyi14}

Without magnetic field vortices with opposite polarities ($p=\pm1$) are energetically equivalent. Under the action of a transversal dc magnetic field the preferable polarity of the vortex coinsides with the field direction (light vortex). The vortex with opposite polarity (heavy vortex) turns into so-called cone phase. Its structure is well-known:\cite{Kosevich83,Ivanov95b} While the in-plane magnetization distribution $\phi$ of the cone phase coincides with \eqref{eq:vortex-phi}, the out-of-plane component is deformed in the following way. The heavy vortex becomes narrower with the typical core width $w_\text{ch}(h) \propto \sqrt{1-|h|}$ in contrast to the light vortex which becomes broader with $w_\text{cl}(h) \propto 1/\sqrt{1-|h|}$.\cite{Ivanov95b}

Instead of the exponentially localized structure \eqref{eq:vortex-m4h=0} without field, the vortex profile has algebraic decay in the cone phase. Far from the origin it is characterized by the asymptote $m=h+h\ell^2/r^2$.\cite{Kosevich83,Ivanov95b} For definiteness we suppose that the vortex polarity $p=+1$ and negative field intensity $h < 0$ having in mind switching phenomenon.

With the field intensity increasing the heavy vortex loses its stability. \cite{Ivanov02} It is important to stress that in spite of the fact that the light vortex is energetically preferable, the heavy vortex can not perform switching in scope of continuous theory due to infinite barrier which separates opposite polarized vortices. However the barrier becomes finite in the \emph{discrete} spin lattice \cite{Wysin94} and the reversal process can occur\cite{Gaididei99,Gaididei00}. The simple picture of the vortex core switching provides the core model, originally introduced by \citet{Wysin94} for the vortex instability phenomenon. Because of the crucial role of exchange interaction inside the vortex core for the switching mechanism, the core model, provides good qualitative description of the switching process. \cite{Pylypovskyi13b,
Pylypovskyi13e} Let us mention one more possibility, the so called cutoff model:\cite{Caputo07} modeling the discreetness effects by cutoff parameter it is possible to describe the vortex polarity switching under the action of spin-polarized current\cite{Caputo07}, by dc\cite{Kravchuk07a} and ac\cite{Pylypovskyi13b} magnetic fields.

\section{Spin-lattice simulations of the vortex polarity switching}
\label{sec:swi}

The main purpose of the current research is to describe the fine structure of the vortex core switching process. As we discussed above the switching is possible only in the discrete lattice. There appear serious difficulties in modeling of this process: the standard micromagnetic simulators consider the numerically discretized Landau-Lifshitz equation, which is valid in continuum theory. Such approach becomes mesh-dependent within the continuum description of micromagnetism.\cite{Thiaville03} The switching is known to be accompanied by the creation of micromagnetic singularities, Bloch points, which can not be described adequately within the continuum limit.\cite{Thiaville03,Pylypovskyi12,Andreas14}

In order to overcome difficulties of continuous approach, we perform spin-lattice modeling  using in-house developed spin-lattice simulator \textsf{SLaSi}. \cite{slasi} Numerically we solve Landau-Lifshitz-Gilbert equations \eqref{eq:LLG} for the disk-shaped system. We consider two sets of parameters: disk diameter $2R=149a$, thickness $L=49a$ and the magnetic length $\ell=14a$ (disks A), $2R=99a$ and $\ell=5a$ with the thickness $L=(N-1)a$, where $N$ is the number of lattice sites along disk axis with $N=\overline{11,50}$ (disks B). Chosen diameters allow vortex to be both in the center of the sample and between lattice sites. Other parameters are following: the initial vortex polarity $p = 1$, the Gilbert damping constant $\varepsilon =0.5$ corresponds to the overdamped regime. The surface anisotropy is chosen $\varkappa_\text{ES}=0.5$ and $\varkappa_\text{EN}=-0.5$, and the field intensity varies in the range $h\in(-0.83;0)$.
The lattice planes are chosen parallel to the face surfaces of the samples.

\subsection{Magnets with bulk anisotropy only: $K_s = 0$}
\label{sec:homogeneous}

\begin{figure*}
\begin{center}
\begin{tikzpicture}
\node at (-7.6,0) {\includegraphics[width=0.14\linewidth]{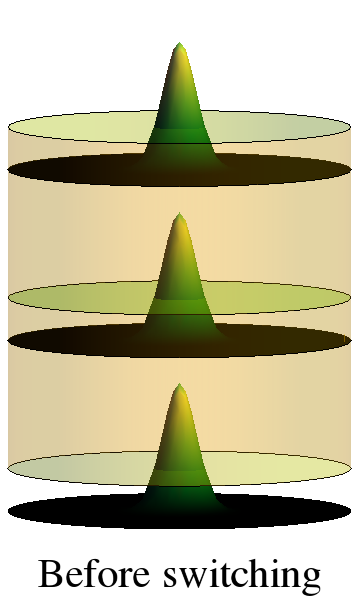}};
\node at (0,0) {\includegraphics[width=0.69\linewidth]{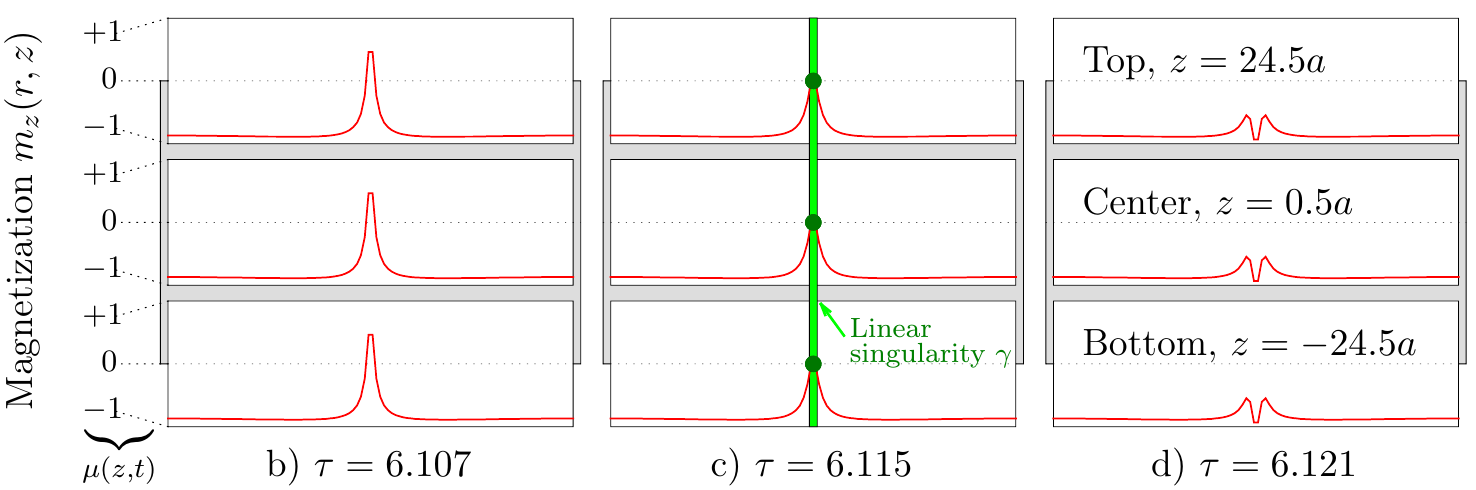}};
\node at (7.6,0) {\includegraphics[width=0.14\linewidth]{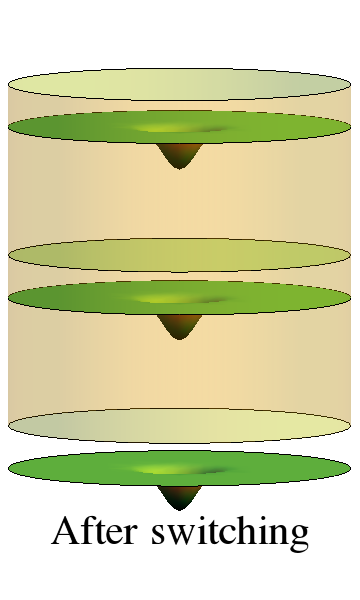}};
\node at (-8.6,1.8) {a)};
\node at (6.6,1.8) {e)};
\end{tikzpicture}
\end{center}
\caption{(Color online) Spatio-temporal picture of the vortex core switching for the magnet with bulk anisotropy only, the simulation data (center). Left (a) and right (e) panels: schematic of the vortex profiles in three different horizontal cross-sections. During the switching process the vortex polarity flips simultaneously in all lattice planes. The gray rectangle on background indicates the XZ cross-section of the sample and the white boxes show the spatial profile of the $z$-component of magnetization $m_z(r,z)$. The magnetization distribution is shown in three different lattice cross-sections ($z_\text{t}=24.5a$, $z_\text{c}=0.5a$ and $z_\text{b} = -24.5a$) at three different time moments: b) before the switching, $\tau = 6.107$, c) at the switching moment, $\tau = 6.115$, and d) just after the switching, $\tau = 6.121$. The switching is accompanied by the linear singularity creation, see the green line. Parameters:
disk~A, $h=-0.83$.}

\label{fig:3d-data}
\end{figure*}

Using the vortex distribution \eqref{eq:vortex} as initial one for different field intensities $h<0$, we relax numerically the system to the static configuration, which is adopted to the lattice. Starting from weak fields, we increase adiabatically its absolute value up to the critical value $h_c$. For stronger field intensities, $|h|\ge |h_c|$, the switching occurs.

We model the switching process for the vortex in the disk~A. Numerically we found that $|h_c|\approx0.83$. The heavy vortex is relaxed in the field with intensity $h=-0.82$. It demonstrates the narrowing of the core and shifting of the background far from the origin described above. After that we applied the field with intensity $h=-0.83$ which results in the polarity switching. The heavy vortex losses its stability: the four central magnetic moments in the lattice planes inside the vortex core rapidly switch their direction to the opposite one. Another magnetic moments relax to the light vortex distribution with a smaller speed. In general, the background of the vortex remains the same, all dynamics is observed only in the core region.

Essentially, the reversal process occurs simultaneously in all lattice planes, hence the switching scenario is fully two dimensional one. The temporal evolution during the switching is shown in the Fig.~\ref{fig:3d-data}. Schematics (a) and (e) represent the 3D magnetization structure in the sample before and after switching, respectively: vortex state within three different horizontal cross-section (top, center and bottom surfaces). Central figures (b)--(d) show the magnetization distribution within the different cross-sections in three different time moments: before the switching ($\tau = 6.107$), at the switching moment ($\tau = 6.115$), and just after the switching ($\tau = 6.121$). During the switching process, the linear singularity $\vec{m}(\vec{\gamma })=0$ is created, see the solid green vertical line $\vec{\gamma }$ in Fig.~\ref{fig:3d-data}.\footnote{Because vortex center at $r = 0$ lies between lattice sites, the length of spins in simulations remains constant.}

\begin{figure}
\begin{center}
\includegraphics[width=\columnwidth]{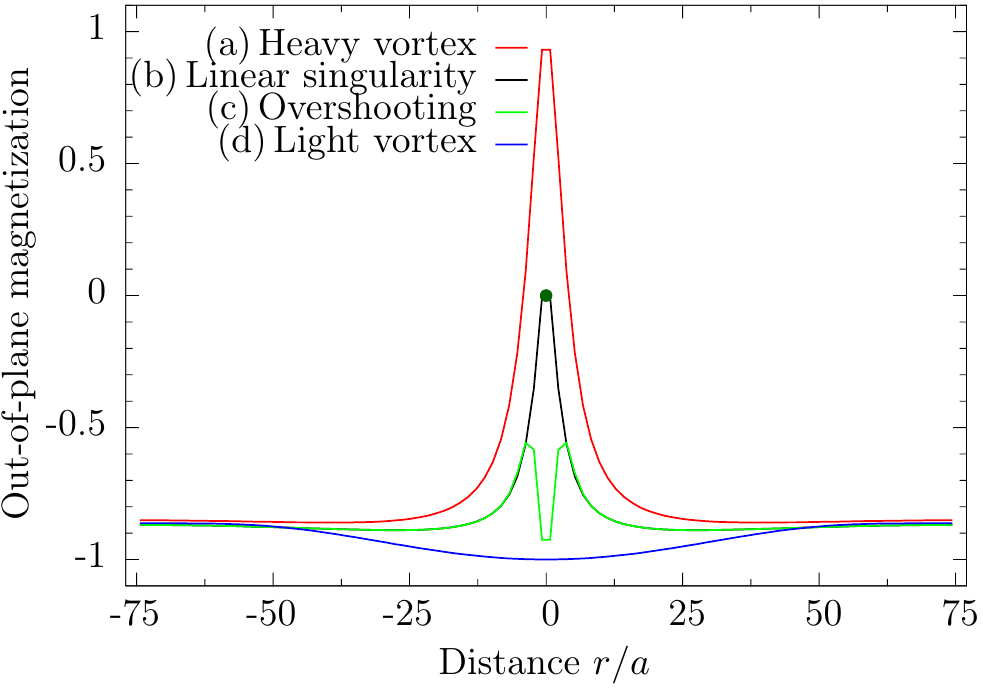}
\end{center}
\caption{(Color online) Vortex polarity reversal in the sequential time moments: (a) Initial state of the heavy vortex before switching; (b) Dynamics polarity in the center decreases due to discreetness, linear singularity appears; (c) Central magnetic moments reverces faster than other ones, overshooting appears; (d) Final state of the light vortex.
}
\label{fig:swi2Dshape}
\end{figure}

In detailes the polarity switching occurs by the following steps, shown in Fig.~\ref{fig:swi2Dshape}. To consider the spin-lattice system we replace the vortex polarity $p = \pm 1$ by a dynamical polarity $\mu=\mu (z,t)$ which represents the actual amplitude of magnetization $z$-component in the vortex core center. The magnetic moments in the central part of the initial heavy vortex core are approximately perpendicular to the plane and one can say that the dynamical polarity is $\mu \approx 1$, see curve (a) in Fig.~\ref{fig:swi2Dshape}. When the applied field becomes stronger than the switching threshold, $\mu $ rapidly decreases and changes its sign passing through $\mu = 0$. One can say about creation of a linear singularity in this moment, see curve (b) in Fig.~\ref{fig:swi2Dshape}. The dynamics of the central magnetic moments is going faster than outer ones. Central moments rapidly flip their direction from $\mu \approx 1$ to
$\mu \approx -1$, and the overshooting in the vortex profile appears, see curve (c) in Fig.~\ref{fig:swi2Dshape}. The
final state is the light vortex, see curve (d) in Fig.~\ref{fig:swi2Dshape}.

It should be noted that during the whole complicated dynamical process of the switching the system remains uniform along the $z$-axis: the numerical difference for different lattice planes is not larger than $10^{-6}$ for $\vec m_{\vec n}$ (corresponds to the accuracy of the saved data) during relaxation in the external field and polarity reversal. The same scenario takes place in the case of non adiabatically applied external field.

\subsection{Magnets with the surface anisotropy}
\label{sec:heisenberg-with-sa}

\begin{figure*}
\begin{subfigure}[b]{0.45\textwidth}
\begin{tikzpicture}
\node at (0,0) {\includegraphics{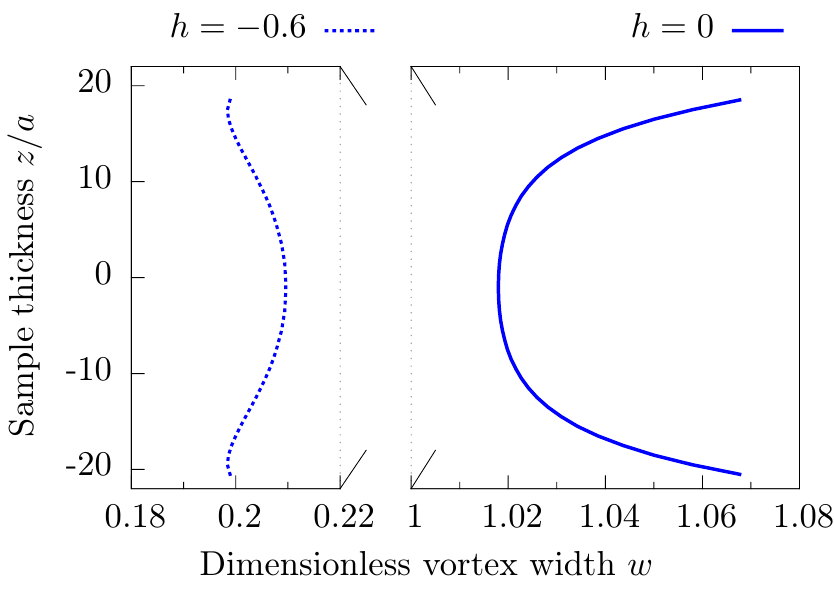}};

\node (pillowh0) at (3,4) {\includegraphics[width=15mm]{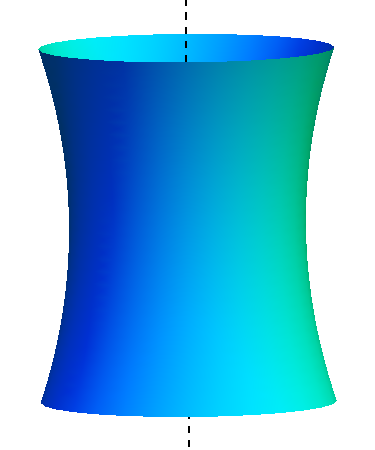}};
\node (pillowh) at (-3,4) {\includegraphics[width=11mm]{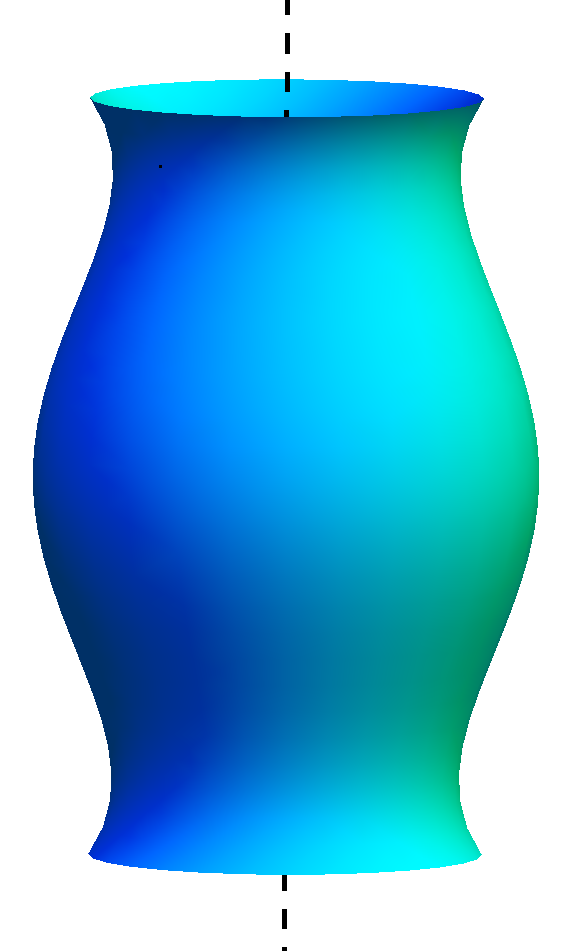}};
\draw[->, double=white, draw=gray] (pillowh0.west) to [out=180, in=0] node [sloped, above] {$|h|$ grows}  (pillowh.east);
\node at (2,3.5) {(c)};
\node at (-2,3.5) {(d)};
\end{tikzpicture}
\caption{EN surface anisotropy, $\varkappa=-0.5$}
\end{subfigure}
\hfill
\begin{subfigure}[b]{0.45\textwidth}
\begin{tikzpicture}
\node at (0,0) {\includegraphics{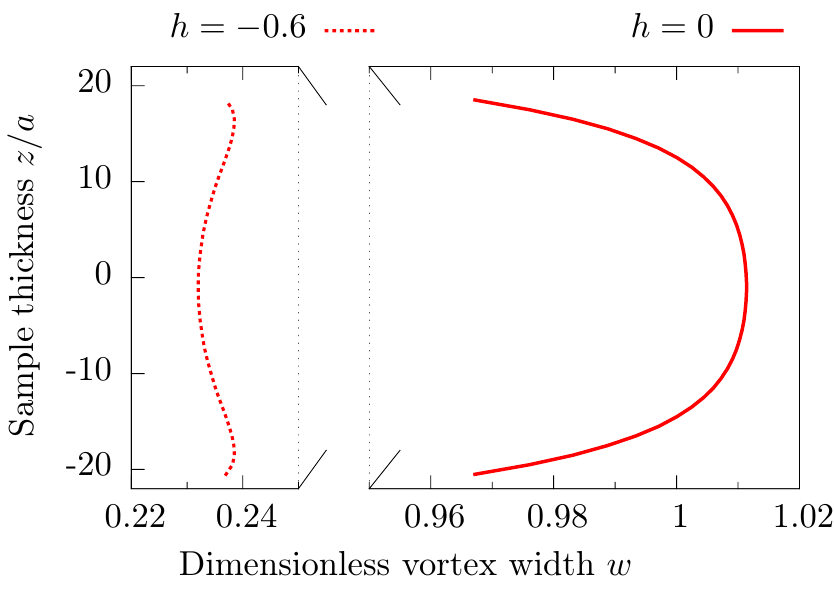}};

\node (barrelh0) at (3,4) {\includegraphics[width=15mm]{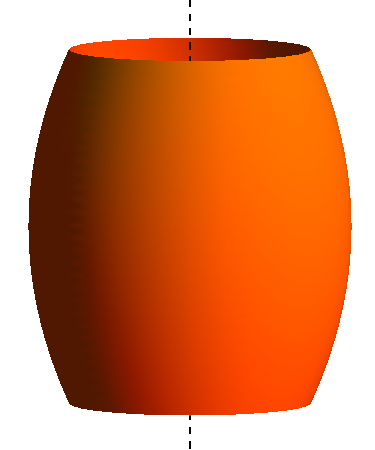}};
\node (barrelh) at (-3,4) {\includegraphics[width=11mm]{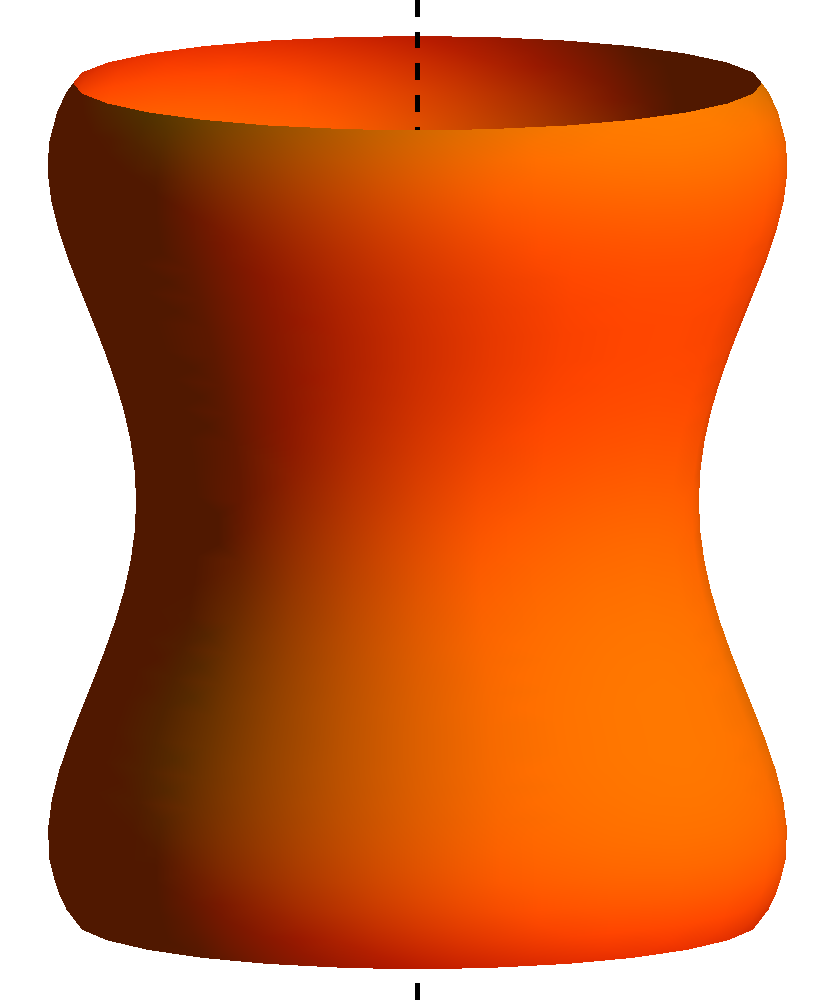}};
\draw[->, double=white, draw=gray] (barrelh0.west) to [out=180, in=0] node [sloped, above] {$|h|$ grows}  (barrelh.east);
\node at (2,3.5) {(e)};
\node at (-2,3.5) {(f)};
\end{tikzpicture}
\caption{ES surface anisotropy, $\varkappa=0.5$}
\end{subfigure}

\caption{(Color online) The reduced vortex core width $w$ as a function of the sample thickness coordinate $z$ for the disk~B with thickness $L=39a$. Without field the vortex core has the pillow shape for the EN case ($\varkappa = -0.5$), see schematic~(c), and the barrel one for the ES case ($\varkappa = 0.5$), see schematic~(e). Under the action of the field close to switching one ($h = -0.6$) there is an opposite behavior: the vortex core profile becomes a deformed barrel one for the EN case, see schematic~(d), and a deformed pillow one for the ES case, see schematic~(f). Arrows show transformation under the action of external field $h$ which is increased by modulus.}
\label{fig:w-via-z-field}
\end{figure*}

\begin{figure*}
\begin{center}
\includegraphics[width=\linewidth]{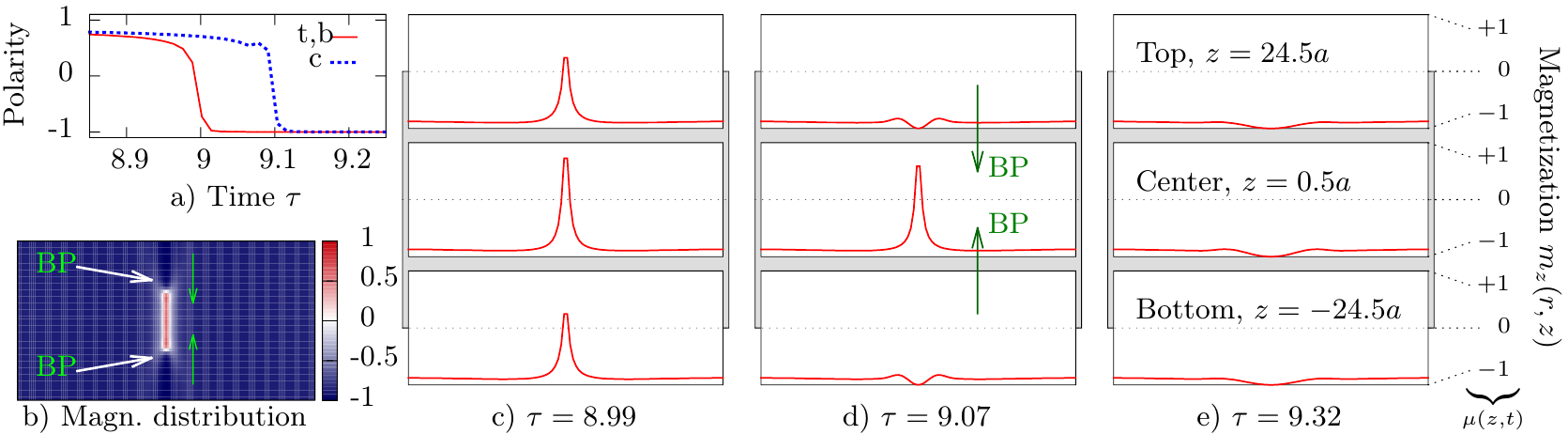}
\end{center}
\caption{(Color online) Switching of the vortex polarity for 3D Heisenberg magnets with EN surface anisotropy (disk~A, $\varkappa = -0.5$, simulations data). a) Polarity dynamics for vortices with $z=24.5a$ and $z=-24.5a$ (solid curves which coincide), $z=0.5a$ (dashed curve). b) Magnetization distribution in the x-z plane near the center of the sample at $\tau=9.07$. Color indicates the $m$ sign, thick arrows show the positions of the Bloch points, thin vertical arrows indicate the direction of motion for the bottom and top Bloch points. c--e) Polarity switching in different x-y planes, according to curves on a). Green arrows show the direction of the Bloch points (BP) motion.
}
\label{fig:EN-data}
\end{figure*}

\begin{figure*}
\begin{center}
\includegraphics[width=\linewidth]{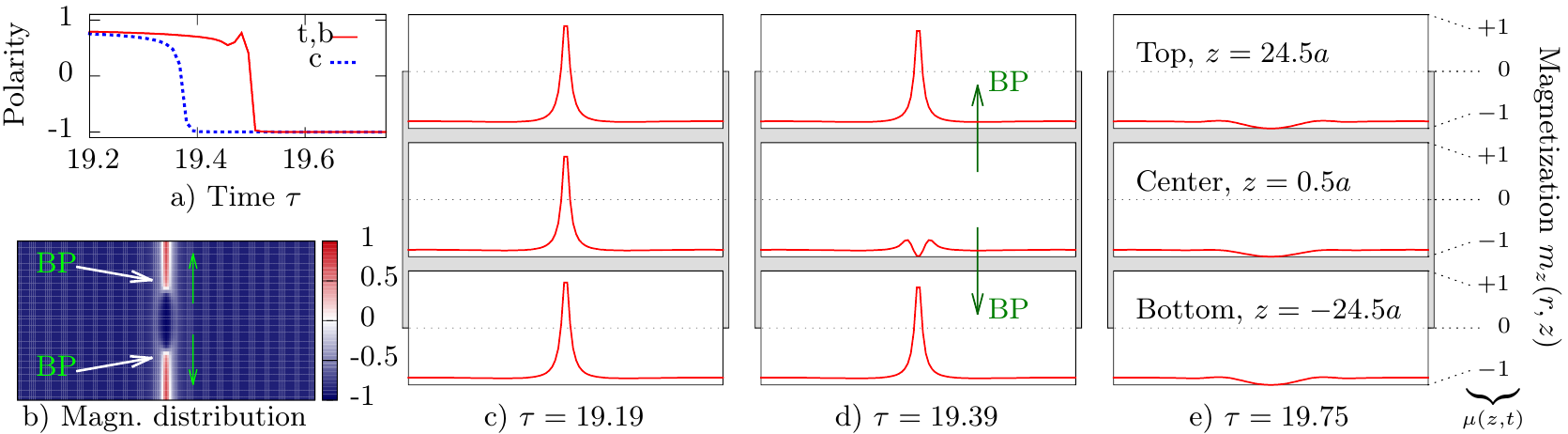}
\end{center}
\caption{(Color online) Switching of the vortex polarity for 3D Heisenberg magnets with additional ES surface anisotropy ($\varkappa = 0.5$), the simulations data. All notations and other parameters are the same as in Fig.~\ref{fig:EN-data}.
}
\label{fig:ES-data}
\end{figure*}

Let us study the role of the surface anisotropy in the vortex statics and dynamics, especially in the vortex switching phenomenon. Very recently we have studied how the surface anisotropy effects on the vortex core shape without magnetic field: there appears the pillow- and the barrel-deformation of the core for the ES and EN anisotropies, respectively. \cite{Pylypovskyi14} Qualitatively, the vortex core width inside the sample volume is determined by the magnetic length, while the core width on a surface layer is characterized  by effective magnetic length\cite{Pylypovskyi14}
\begin{equation} \label{eq:ell-eff}
\ell_{\text{eff}}=\frac{\ell}{\sqrt{1+\varkappa }}.
\end{equation}
In the case of ES surface anisotropy ($\varkappa>0$), an effective magnetic length $\ell_{\text{eff}}<\ell$. Therefore the vortex core becomes more narrow near the face surfaces. This corresponds to the barrel-shaped profile. In the same way, EN surface anisotropy ($\varkappa<0$) results in the pillow-shaped profile because $\ell_{\text{eff}}>\ell$, see Fig.~\ref{fig:w-via-z-field} and schematics~(c) and~(e) therein.

In weak fields (far from the switching threshold) the vortex core looks qualitatively the same: barrel-shaped profile for the ES case and the pillow-shaped one for the EN case, while core becomes more narrow in agreement with above mentioned estimations.

Under the action of intermediate field the vortex core profile deforms. The further increase of the field intensity (up to the switching threshold) drastically changes the vortex core shape. In the ES case there appears a deformed pillow core instead of original barrel one, see the Fig.~\ref{fig:w-via-z-field} and schematic~(f) therein. In the same way in the EN case one has the deformed barrel core profile instead of original pillow one, see the Fig.~\ref{fig:w-via-z-field} and schematic~(d) therein.
Numerically the vortex core width is computed as the $w_0(z)$ such that $m(w_0) = 0.5\left[m(0) + m(R)\right]$.

The inhomogeneous shape of the vortex core profile is a key point for understanding the Bloch point mediated switching scenario. Let us start with the case of EN surface anisotropy. In this case the vortex core profile becomes deformed barrel-shaped, see the Fig.~\ref{fig:w-via-z-field} with schematic~(d). With the field increasing both bending points shift outside to the ends of the Bloch line and they become closer to the face surfaces when field approaches to the switching one. This means that in a strong enough field the vortex profile near the face surface is more narrow than in the bulk.

This bulk deformation of the vortex core results in the inhomogeneous vortex polarity switching: the Bloch line breaks in its more narrow parts, \emph{i. e.} on the face surfaces. Each break of the Bloch line corresponds to the Bloch point. Therefore in case of EN surface anisotropy two Bloch points are nucleated simultaneously on the face surfaces. During the reversal process they move in opposite direction to each other and finally
annihilate at
the sample's center. The similar behaviour occurs in the case of ES surface anisotropy: the narrowest part of the vortex core is situated in the center and a pair of Bloch points is nucleated in the center of the sample.

\begin{figure*}
\begin{subfigure}[b]{0.48\textwidth}
\includegraphics[width=\columnwidth]{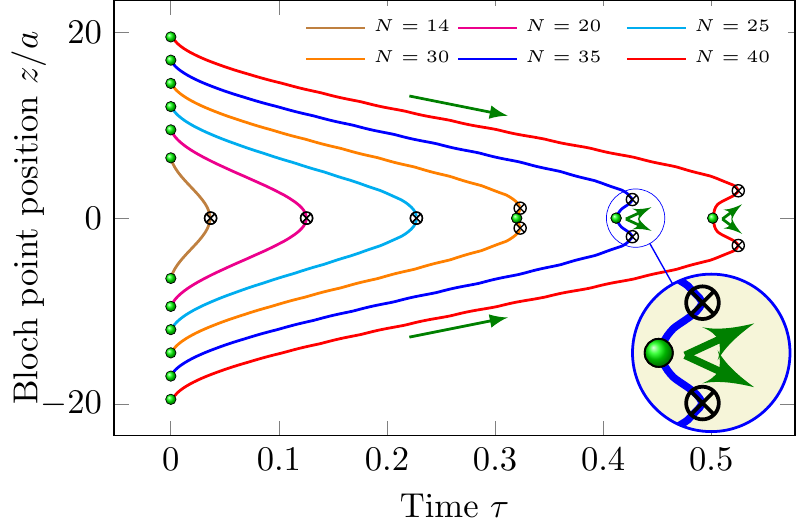}
\caption{EN surface anisotropy, $\varkappa=-0.5$}
\end{subfigure}
\hfill
\begin{subfigure}[b]{0.48\textwidth}
\includegraphics[width=\columnwidth]{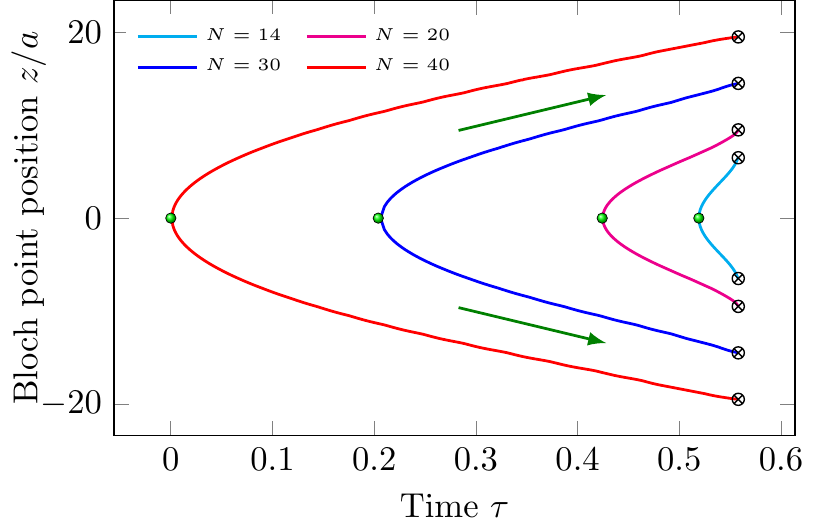}
\caption{ES surface anisotropy, $\varkappa=0.5$}
\end{subfigure}
\caption{(Color online) Bloch point position as function of time for samples with different thickness. All curves are matched at $\tau=0$ for EN surface anisotropy and at $\tau=0.548$ for ES surface anisotropy. Green circles indicate Bloch points creation event and crosses indicate annihilation event. For $N\ge 30$ additional pair of Bloch points is nucleated shortly before annihilation. Arrows indicate direction of motion of Bloch points. Parameters: disk~B, $\varkappa=-0.5$, $h = -0.8$.
}
\label{fig:bp-motion}
\end{figure*}

The schematics of the 3D polarity reversal accompanied by Bloch points motion for EN surface anisotropy is shown in Fig.~\ref{fig:EN-data}. The simulation data are shown for the disk~A under the action of the external field $h=-0.85$ applied to the sample relaxed in $h = -0.8$. Bloch points always separate vortices with opposite polarities. Therefore their existence can be easily detected by observing of the vortex polarities on the different lattice planes: face surfaces and in the center of the sample, see Fig.~\ref{fig:EN-data}a. The temporal evolution of both Bloch points has the reflection symmetry with respect to $z=0$ plane: curves which show polarity dynamics of top and bottom vortices coincide (solid curve). The switching of the central vortex occurs with delay (dashed curve). In thicker samples more pairs of Bloch points can be nucleated inside the sample volume, see below (also see inset in Fig.~\ref{fig:bp-motion}).

The axial cut of the lattice during polarity switching is shown in the Fig.~\ref{fig:EN-data}b: the vortex state is already reversed ($\mu \approx -1$) near the top and bottom surfaces, while its polarity is still directed up ($\mu \approx +1$) in the sample's center. There are two Bloch points in the system: their positions are shown by thick white arrows and the direction of motion is shown by thin green arrows. The vortex profiles for different times during polarity switching process are shown in the Fig.~\ref{fig:EN-data}c--e.

Let us discuss now the case of ES surface anisotropy, see Fig.~\ref{fig:ES-data} (notations are the same as in Fig.~\ref{fig:EN-data}). Here we apply the external field $h=-0.85$ to the sample relaxed in $h = -0.8$. It results in the deformed pillow-shaped vortex core profile. That is why the vortex core becomes more narrow in the middle of the sample, hence the Bloch line breaks in the middle and the switching process starts inside the sample. There are two Bloch points which born in the sample center. During the switching process they move along $z$-axis in opposite direction to face surfaces.

Numerically we also analyzed the case of the exchange anisotropy; qualitatively the mechanism is very similar to the EN surface anisotropy, see Appendix \ref{sec:Heisenberg-SAex-numerics} for details.

To gain insight into the temporal evolution of the switching process we performed a set of simulations for fixed diameter of the sample (disk~B) and varying thicknesses ($N=\overline{11,40}$). The anistropy is $\varkappa=-0.5$. We apply the field $h=-0.8$\footnote{Step-wise field turning-on decreases the switching threshold.} to the vortex state relaxed without field ($h=0$). Similarly to the previous case, the inhomogeneous polarity reversal is observed. In all simulations the reversal is accompanied by the nucleation of a pair of Bloch points near face surfaces, their further dynamics inside the magnet and the final annihilation at the point of contact in the middle. The temporal dynamics of the Bloch points for different thicknesses is shown in Fig.~\ref{fig:bp-motion}. We determine the position of Bloch points as a cross-sections of three isosurfaces $m_\alpha (\vec r) = 0$ with $\alpha =x,y,z$.\cite{Hertel06} Each surface $m_\alpha (
\vec r)$ is computed as interpolated functions for magnetization components based on discrete values of $\vec m_{\vec n}$.

A new feature appears for relatively thick samples ($N\ge 30$): During the switching process, additional pair of Bloch points is nucleated in the sample center. Two points of this pair repel each other moving in $z$-direction out from the center. Finally the new born Bloch points annihilate with originally nucleated ones, see the inset in Fig.~\ref{fig:bp-motion}. Note that for the weaker field $h=-0.7$ the additional Bloch points are not nucleated. One can conclude that there appears instability of the Bloch line in long enough samples, which causes the breaking of the Bloch line inside the sample and nucleation of additional Bloch points. Such a picture is similar to the temporal formation of micromagnetic drops in long nanocylinders.\cite{Hertel04a}

\begin{figure}
\begin{center}
\includegraphics[width=\columnwidth]{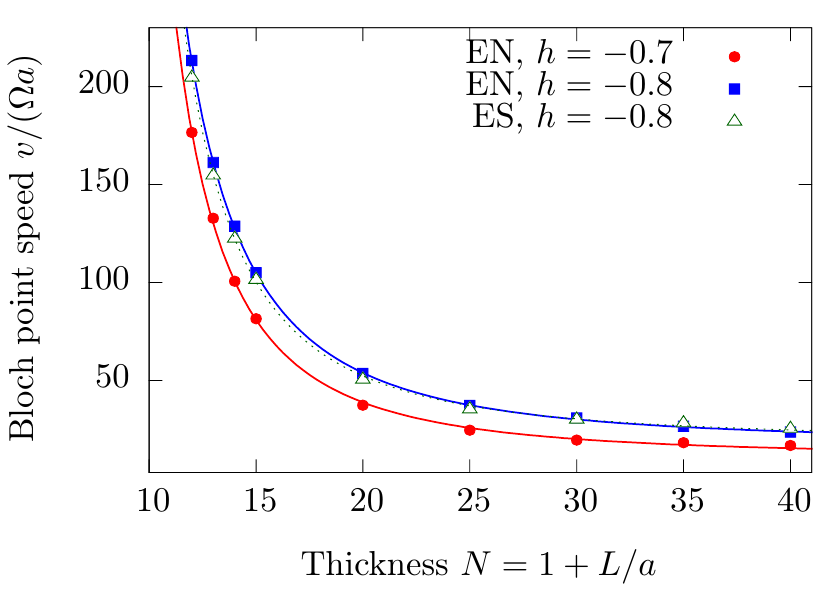}
\end{center}
\caption{(Color online) Bloch point speed as function of sample thickness in units of lattice sites along disk axis for the EN and ES surface anisotropies. Parameters are the same as in Fig.~\ref{fig:bp-motion}.
}
\label{fig:bp-speed}
\end{figure}

Numerically we also studied how the speed of the Bloch point depends on the sample thickness. One can see from the Fig.~\ref{fig:bp-motion} that the Bloch point  motion is almost a steady-state one except the moments of nucleation and annihilation. In this way we computed the speed of Bloch points $v$ in a wide range of the sample thickness, see Fig.~\ref{fig:bp-speed}. The Bloch point speed rapidly decreases in thick samples and tends to some constant value not dependend on the sample thickness and surface anisotropy type.

\section{Description of the polarity switching process}
\label{sec:wcm}

To describe analytically the observed vortex polarity switching process we propose a \emph{wired core model}. In spite of simplicity this model is in a good qualitative agreement with full-scale simulations. It generalizes the discrete reduced vortex core model initially proposed by \citet{Wysin94} for the vortex instability phenomenon and later extended for the description of the polarity switching in the two-dimensional case. \cite{Gaididei99, Gaididei00, Zagorodny03, Kovalev03, Pylypovskyi13e} In the current study we extend the model for 3D systems. A wired core model allows one to take into account processes of nucleation-annihilation of point singularities during the vortex switching.

We start with the discrete Hamiltonian \eqref{eq:H-total}. The anisotropy term \eqref{eq:H-EP+SA} can be written in the following form:
\begin{equation} \label{eq:H-EP+SA-2}
\begin{split}
\mathcal{H}^{\text{an}} &= \frac{\mathcal{S}^2}{2} \sum_{\vec n} K_n m_{\vec{n}}^2, \\
K_n &= \begin{cases}
K,     & \text{for $n=\overline{2,N-1}$}\\
K+K_s, & \text{for $n=1$ and $n=N$}.
\end{cases}
\end{split}
\end{equation}
Here $\vec n=(n_x,n_y,n)$ and $n$ enumerates the planes along $z$ direction.

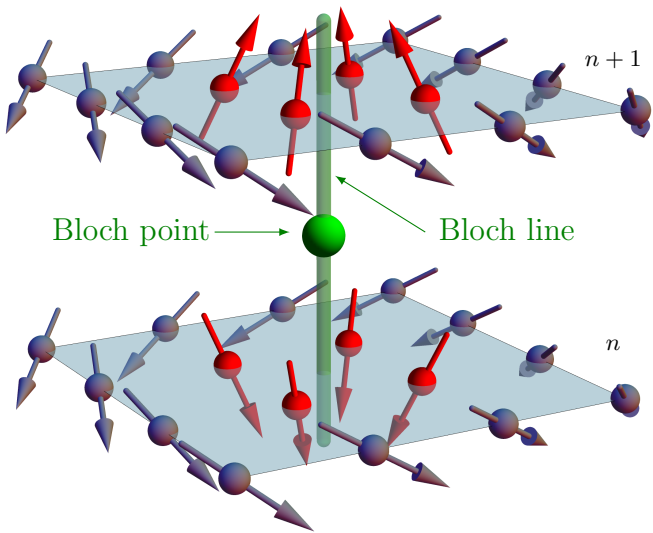
\begin{figure}
\begin{center}
\begin{tikzpicture}

\node at (0,0) {\includegraphics[width=\columnwidth]{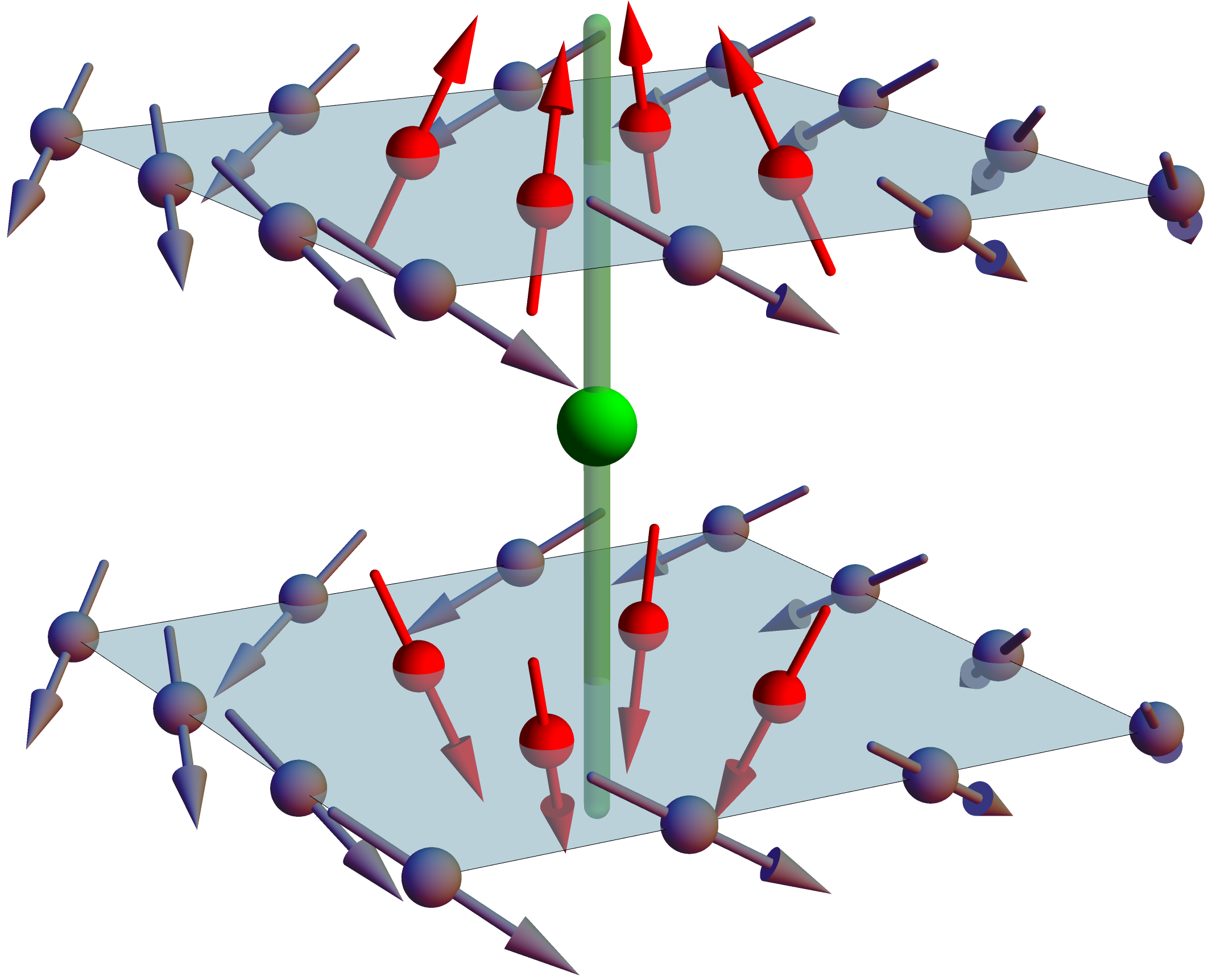}};

\node at (3.8,-1.) {$n$};
\node at (3.8,2.8) {$n+1$};
\draw[->,green!50!black] (-1.5,0.5) -- +(1,0) node[xshift=-60pt,yshift=0pt] {\large Bloch point};
\draw[->,green!50!black] (1.3,0.5) -- +(-1.2,0.75) node[xshift=65pt,yshift=-20pt] {\large Bloch line};
\end{tikzpicture}
\end{center}
\caption{(Color online) Schematic of the wired core model for $n$-th and $(n+1)$-th planes: Red arrows  indicate free magnetic moments (Eq.~\eqref{eq:freeSpins}) and gray ones indicate fixed ones (Eq.~\eqref{eq:fixedSpins}). Layers are exchange-coupled with exchange integral $J$ and the each plane is characterized by own anisotropy constant $K_n>0$, see Appendix~\ref{app:core} for details. Bloch point position is marked by the green sphere.}
\label{fig:core-schematic}
\end{figure}

We suppose that in each $n$-th plane there exist only four ``free'' magnetic moments, all other moments are ``frozen'' into the background, see Fig.~\ref{fig:core-schematic}. The 3D vortex core is considered in our model as a wire of free magnetic moments.

In original core model\cite{Wysin94} the discrete core magnetization is matched with the continuous magnetization of the background (the ring), where the magnetization is fixed in the pure planar vortex configuration. The inner radius $r_\text{in}$ of the ring should be chosen from the condition that the ground state is a pure planar vortex, which fulfills when $r_\text{in} \ge 0.3 \ell$.\cite{Kravchuk07} The situation drastically changes under the field action, the planar vortex can never be a solution of the ring wire in the field presence. Besides one has to take into account $z$--dependence of the magnetization due to the surface anisotropy in the way similar to \eqref{eq:m0-z}. Finally, we get the background magnetization distribution $m_r(r, z)$, see Appendix~\ref{app:ring} for details:

\begin{subequations} \label{eq:freefixedSpins}
\begin{equation} \label{eq:fixedSpins}
\begin{split}
m_{(n_x,n_y,n)} &= m_b(z_n) = h m_r(r_2, z_n),\\
\phi_{(n_x,n_y,n)} &= \chi+ \frac{\pi }{2},\qquad |n_x|+|n_y| > 2,
\end{split}
\end{equation}
where $r_2$ is the radius of the second coordination sphere, function $m_r$ is defined by Eq.~\eqref{eq:mrring}, $z_n$ is a $z$-coordinate of $n$-th lattice plane.  For the numerical investigation of the wired core model we use $r_2 \approx 0.35\ell$.

All ``free'' moments come from the ring of principal sites, they form the vortex core. Due to the axial symmetry, all free moments have the same out-of-plane component $\mu _n$ and the same in-plane phase $\psi$ which describes the deviation from the vortex distribution \eqref{eq:fixedSpins}:
\begin{equation} \label{eq:freeSpins}
\begin{split}
m_{(n_x,n_y,n)} &= \mu_n(t),\\
\phi _{(n_x,n_y,n)} &= \chi + \frac{\pi }{2} + \psi _n(t),\qquad n_x, n_y=-1, 1.
\end{split}
\end{equation}
\end{subequations}
We consider the \emph{dynamical vortex polarity} $\mu_n(t)$ for each lattice plane, and the in-plane magnetization angle $\psi _n(t)$, which has the meaning of the \emph{turning phase}, as two collective variables. In this manner we generalize the 2D approach, which we recently proposed in Ref.~\onlinecite{Pylypovskyi13e}. Note that the dynamical vortex polarity is mutually connected to the effective vortex width in the full-scale model.

Now by incorporating the wired core Ansatz \eqref{eq:freefixedSpins} into the Hamiltonian \eqref{eq:H-total} with account of the anisotropy term \eqref{eq:H-EP+SA-2}, one can write down the model Hamiltonian, normalized  by $K\mathcal{S}^2$ as follows
\begin{equation} \label{eq:Hwcm1}
\begin{split}
\mathscr H_c = & 4\sum_{n=1}^N \left[\left(1 + \varkappa_n - 2\lambda \right)\frac{\mu_n^2}{2} - \left(h + 2\lambda m_b\right)\mu_n \right. \\
& \left.- \frac{4\lambda }{\sqrt{5}}\sqrt{(1-m_b^2)(1-\mu_n^2)}\cos \psi_n \right] \\
& - 4\lambda \sum_{n=1}^{N-1} \Bigl[ \mu_n\mu_{n+1}  + \sqrt{(1-\mu_n^2)(1-\mu_{n+1}^2)} \\
& \times \cos (\psi_n-\psi_{n+1}) \Bigr].
\end{split}
\end{equation}
Here $\varkappa_n=0$ for $n=\overline{2,N-1}$ and $\varkappa _n=\varkappa $ for $n=1,N$, see Appendix~\ref{app:core} for details. The first sum in \eqref{eq:Hwcm1} describes the intra-plane interaction and the second sum describes the inter-plane interaction.

The equilibrium values of $\mu_n$ and $\psi_n$ can be found analytically for the homogeneous case $\varkappa_n=0$, $n=\overline{1,N}$ and $h=0$:
\begin{equation*}
\mu_\text{hom}(h=0) = \pm \sqrt{1 - \frac{16\lambda^2}{5\left(2\lambda-1\right)^2}},\quad \psi_0(h=0)=0,
\end{equation*}
where $\lambda > 10$ for the out-of-plane vortex. For $h\neq 0$, the dependence $\mu_\text{hom}(h)$ can be found numerically as the solution of the Eq.~\eqref{eq:muhomh}, see Fig.~\ref{fig:polarityInField} in Appendix~\ref{app:core}.

When the field is applied one has to take into account that the equilibrium $z$-components of magnetization are different for the bulk and surface planes. For the fixed magnetic moments the out-of-plane magnetization component is taken equal to $m_b(z)$ according to Eq.~\eqref{eq:mrring}. Assuming that the deviation $x(z)\varkappa/\sqrt{\lambda }\equiv \mu (z) - \mu_\text{hom}$ from the equilibrium value $\mu_\text{hom}$ is small, $|\varkappa/\sqrt{\lambda }|\ll1$, and replacing the Hamiltonian~\eqref{eq:Hwcm1} by its continuum version we obtain the following boundary value problem for $x(z)$:
\begin{equation}\label{eq:wcm-bvp}
\begin{split}
a^2 x''(z) & - A_1 x(z) = A_2(z),\\
\pm a x'(z) & + \frac{\mu_\text{hom}(1-\mu_\text{hom}^2)}{\sqrt{\lambda }}\Bigr|_{z=\pm L/2} = 0,
\end{split}
\end{equation}
where coefficients $A_1$ and $A_2(z)$ are given by Eqns.~\eqref{eq:wcm-coefficients}, see Appendix~\ref{app:core} for details. Function $A_2(z)$ contains only simple harmonics and hyperbolic sine and cosine, so the solution of Eq.~\eqref{eq:wcm-bvp} can be found analytically, nevertheless it is unwieldy and below we present only numerical solution.

\begin{figure}
\begin{center}
\includegraphics[width=\columnwidth]{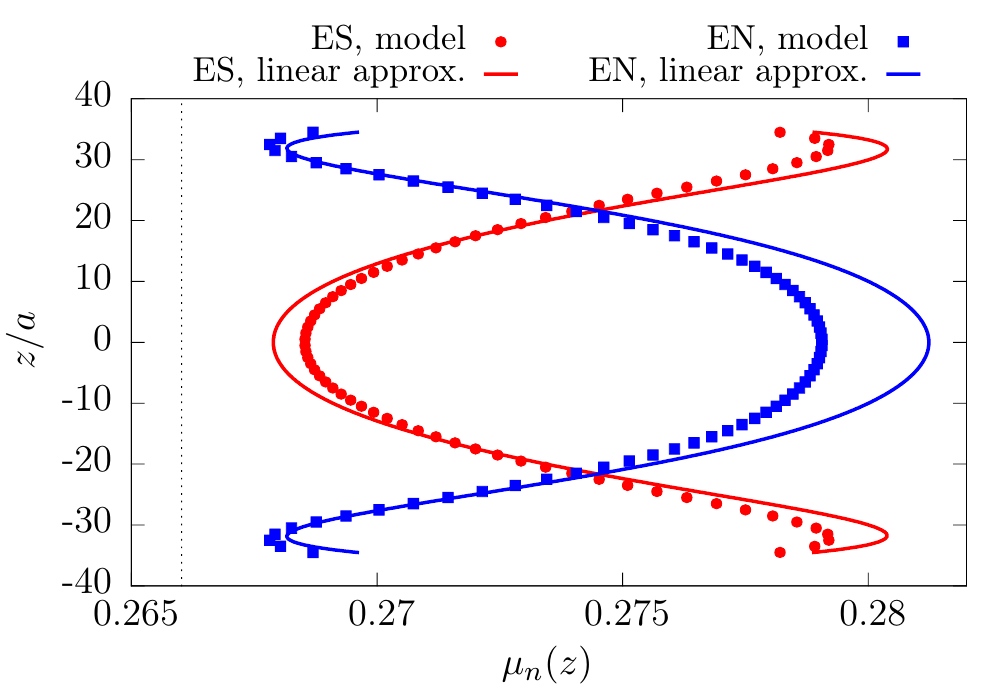}
\end{center}
\caption{(Color online) Dynamical polarity $\mu_n$ for the discrete model~\eqref{eq:Hwcm1} (symbols labeled EN and ES model) and $\mu (z)$ as numerical solution of Eq.~\eqref{eq:wcm-bvp} (solid curves labeled EN and ES linear approx.). Parameters: $|\varkappa| = 0.1$, $\lambda=20$, $N=70$, $h=-0.0009$. Dashed line corresponds to polarity without surface anisotropy.
}
\label{fig:wcm-field}
\end{figure}

\begin{figure*}
\begin{subfigure}[b]{0.48\textwidth}
\includegraphics[width=\columnwidth]{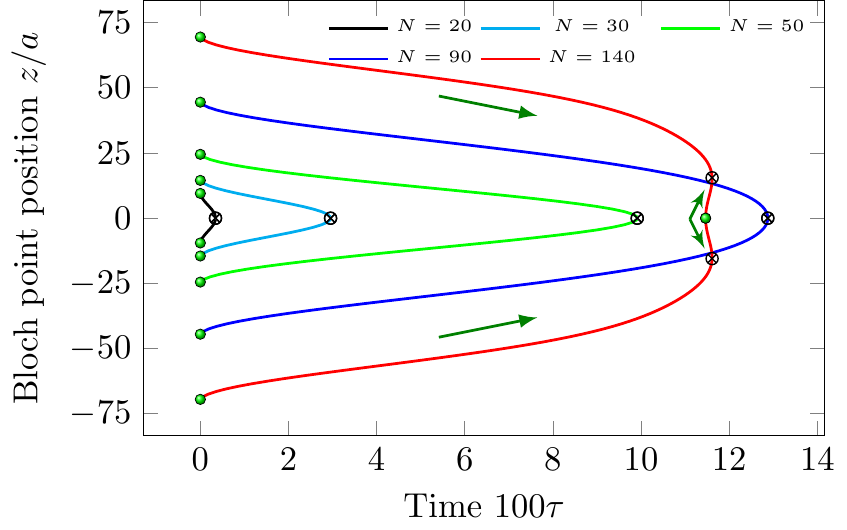}
\caption{EN surface anisotropy, $\varkappa=-0.1$}
\end{subfigure}
\hfill
\begin{subfigure}[b]{0.48\textwidth}
\includegraphics[width=\columnwidth]{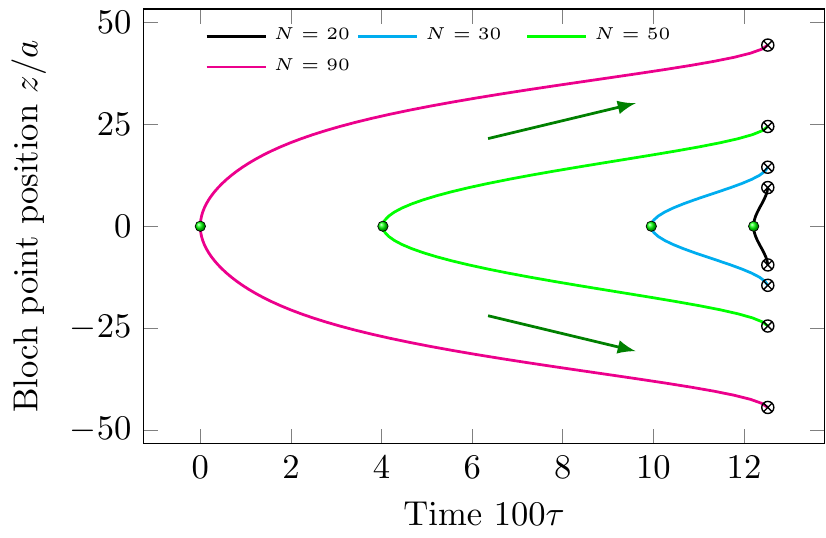}
\caption{ES surface anisotropy, $\varkappa=0.1$}
\end{subfigure}
\caption{(Color online) Bloch point position as function of time for samples with different thickness for the \textit{wired core model}. All curves are matched at $\tau=0$ for EN surface anisotropy and at $\tau=0.125$ for ES surface anisotropy. Green circles indicate Bloch points creation event and crosses indicate annihilation event. For $N = 140$ additional pair of Bloch points is nucleated shortly before annihilation. Arrows indicate direction of motion of Bloch points. $\varepsilon=0.5$, $h=-0.001$, other parameters are the same as in Fig.~\ref{fig:wcm-field}.
}
\label{fig:wcm-motion}
\end{figure*}

The dynamical polarity as numerical solution of Eq.~\eqref{eq:wcm-bvp} for both cases of surface anistoropy is plotted in Fig.~\ref{fig:wcm-field} by solid lines. One can see this result is in a good agreement with a direct solution for Hamiltonian~\eqref{eq:Hwcm1} (symbols correspond to simulations of the system with $|\varkappa|=0.1$, $\lambda=20$ and $N=70$). Numerically we minimize \eqref{eq:Hwcm1} with respect to $\mu_n$ and $\psi_n$ under the action of field $h=-0.0009$. In equilibrium $\psi_n = 0$. The corresponding equilibrium polarity in the homogeneous system $\mu_\text{hom}(h) = 0.266$ is shown by the dashed line.

The dynamical polarity in the wired core model plays a role of the vortex width $w(z)$ in the full-scale simulations. They reproduce the same shapes near the switching field, cf.~Fig.~\ref{fig:w-via-z-field} and~\ref{fig:wcm-field}, and the switching starts in places with minimal $\mu_n$. The deformation of the vortex shape in the wired core model is a result of the inhomogeneous ground state $m_b(z)$ and is highly sensitive to its exact form.

The switching occurs when field increases to $|h|=0.001$. As in the full-scale simulations discussed above the Bloch points nucleate on the surface for EN surface anisotropy, see Fig.~\ref{fig:wcm-motion}a, cf.~Fig.~\ref{fig:bp-motion}a; they propagate from the lattice planes with $n=1$ and $n=N$ to the center of the axis. For a thick enough lattice systems additional Bloch points are nucleated, see curve for $N = 140$ in the Fig.~\ref{fig:wcm-motion}a. Bloch points are nucleated in the center of the system for ES surface anisotropy in the same way as in the full-scale simulations, see Fig.~\ref{fig:wcm-motion}b, cf.~Fig.~\ref{fig:bp-motion}b.

We also studied thickness dependence of the Bloch point speed in the same way as in full-scale simulations, see the Fig.~\ref{fig:wcm-motion} for the field $h = -0.001$ (numerical solution of Eqns~\eqref{eq:llg-wcm} with $\varepsilon = 0.5$. The comparison with the full-scale simulations (cf.~Fig.~\ref{fig:bp-speed}) confirm the fast decay of the Bloch wall speed with increasing of the sample thickness with the further saturation.

\begin{figure}
\begin{center}
\includegraphics[]{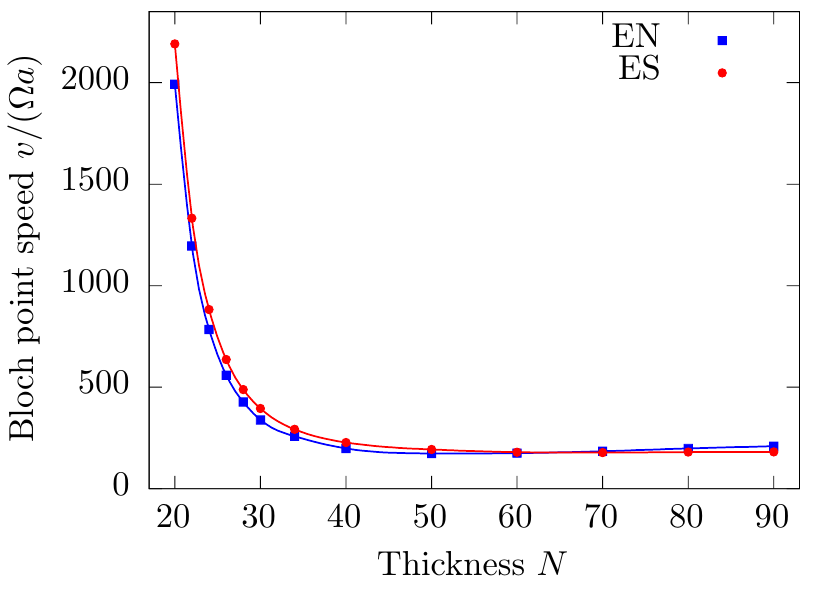}
\end{center}
\caption{(Color online) Bloch point speed as function of sample thickness in the wired core model, cf. Fig.~\ref{fig:bp-speed}. Parameters are the same as in Fig.~\ref{fig:wcm-motion}.
}
\label{fig:wcm-speed}
\end{figure}

\section{Discussion}
\label{sec:discussion}

\begin{figure*}
\begin{subfigure}[b]{0.45\textwidth}
\begin{tikzpicture}
\node at (0,0) {\includegraphics{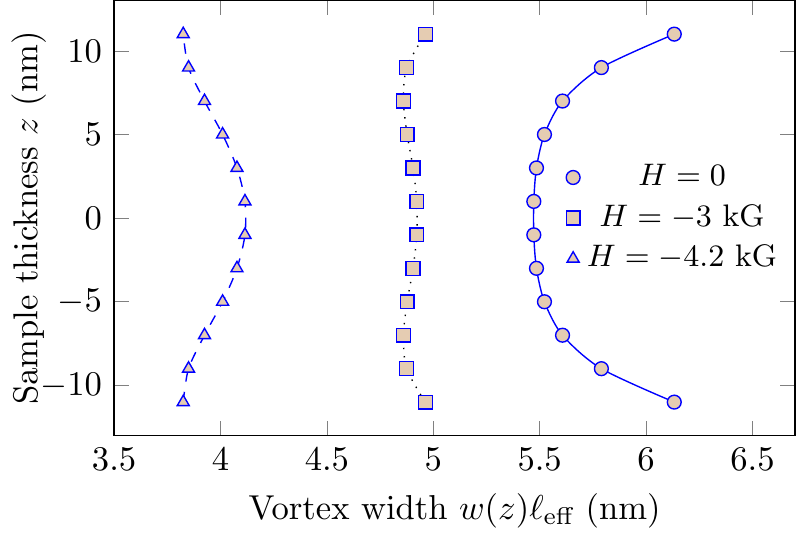}};

\node (pillowh0) at (3,4) {\includegraphics[width=15mm]{pillowh0}};
\node (pillowh) at (-3,4) {\includegraphics[width=11mm]{pillowh}};
\draw[->, double=white, draw=gray] (pillowh0.west) to [out=180, in=0] node [sloped, above] {$|h|$ grows}  (pillowh.east);
\node at (2,3.5) {(c)};
\node at (-2,3.5) {(d)};
\end{tikzpicture}
\caption{EN surface anisotropy, nanodisk of radius 100~nm and thickness 24~nm}
\end{subfigure}
\hfill
\begin{subfigure}[b]{0.45\textwidth}
\begin{tikzpicture}
\node at (0,0) {\includegraphics{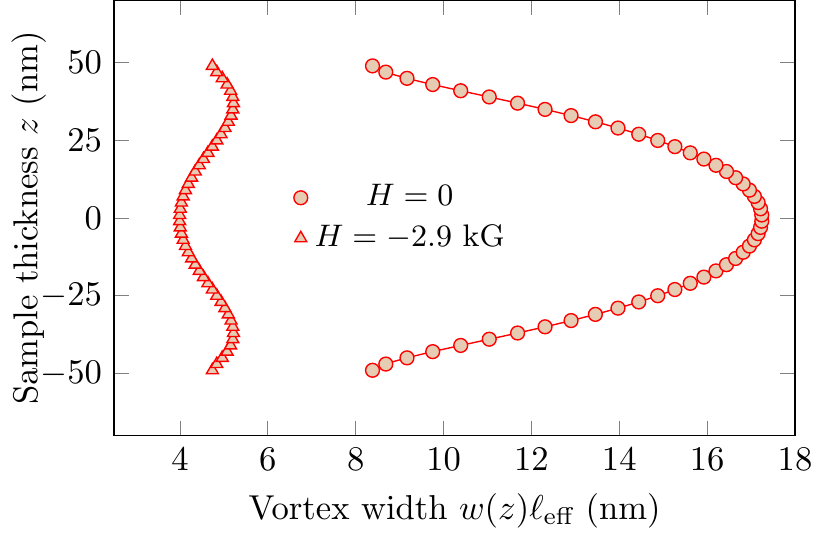}};

\node (barrelh0) at (3,4) {\includegraphics[width=15mm]{barrelh0}};
\node (barrelh) at (-3,4) {\includegraphics[width=11mm]{barrelh}};
\draw[->, double=white, draw=gray] (barrelh0.west) to [out=180, in=0] node [sloped, above] {$|h|$  grows}  (barrelh.east);
\node at (2,3.5) {(e)};
\node at (-2,3.5) {(f)};
\end{tikzpicture}
\caption{Effective ES surface anisotropy induced by dipolar interaction, nanodisk of radius 66~nm and thickness 100~nm}
\end{subfigure}

\caption{(Color online) Simulations with dipolar interaction using \textsf{OOMMF} for EN and effective ES surface anisotropies. Schematics show the change of the vortex profile in the same way as in Fig.~\ref{fig:w-via-z-field} }
\label{fig:oommf}
\end{figure*}

To summarize, we described different mechanisms of the vortex polarity switching in Heisenberg magnets with and without additional surface anisotropy. Under the action of transversal dc magnetic field the vortex core is reversed using the axially-symmetric scenario.\cite{Kravchuk07a, Pylypovskyi13b, Pylypovskyi13e} We conclude that without surface anisotropy the switching occurs uniformly with respect to the thickness $z$-coordinate through the transient linear singularity or the planar vortex. In presence of the surface anisotropy the switching is accompanied  by nucleation of point singularities (Bloch points), their motion and final annihilation. The birthplace of Bloch points and the direction of their motion depends on the type of the surface anisotropy. The complicated vortex dynamics including its switching obtained using full-scale spin-lattice simulations can be described analytically by a wired core model,
which is elaborated in this work.

Qualitatively, the influence of the surface anisotropy can be explained as follows. In terms of the surface anisotropy the magnetization configuration is pinned near the sample boundary. This effective pinning has to compete with exchange interaction, which results in the Robin type boundary condition \eqref{eq:BVP-tot-2}. That is why the homogeneous magnetization distribution $m=h$ is not possible, since it does not satisfy the boundary condition; there appears $m_0(z,h)$ background profile, see \eqref{eq:m0-z}. Similar scenario takes place for the vortex state particle, where the vortex width becomes $z$-dependent, $w=w(z)$. According to \eqref{eq:BVP-tot-2} the sign of the vortex core width gradient depends on the surface anisotropy type: barrel-shaped for the ES surface anisotropy and pillow-shaped for the EN one. \cite{Pylypovskyi14}

Let us sketch the physical picture of the influence of the surface anisotropy on statics and dynamics using the particular case of ES  surface anisotropy. The typical vortex core width $w$ on a surface layer without magnetic field is determined by effective magnetic length $\ell_{\text{eff}}$, see \eqref{eq:ell-eff}.
In ES case, the value of effective magnetic length $\ell_{\text{eff}}<\ell$, and the vortex core becomes more narrow near the face surfaces, which corresponds to the barrel-shaped profile. Under the action of magnetic field the vortex state undergoes several changes. First of all, there appears a cone state instead of the easy-plane one. Apart this the core width for the heavy vortex becomes narrower, $w(h)\propto \sqrt{1-|h|}$. However there is a counteraction of the surface anisotropy, which tries to fix the $z$-gradient of the core width $w(z)$. As a result of this competition, there appears a neck in the vortex core profile, which results in the deformed barrel core instead of original barrel one, Fig.~\ref{fig:w-via-z-field} with schematics~(e) and~(f). This static picture plays a crucial role for the understanding the reversal mechanism. When the field intensity approaches the critical value (switching field), the vortex core width extremely decreases. However this happens
nonhomogeneously: the Bloch line has a bottleneck in its center, hence it breaks during the switching in this place. Namely this point is the birthplace of two Bloch points, which move along the Bloch line repelling each other until they annihilate on the face surfaces of the sample.

In the current study we do not take into account the dipolar interaction, which is of great importance for the real magnets. Specifically, namely the dipolar interaction favors the magnetization curling and causes that the vortex becomes the ground state of magnetically soft nanodisk.\cite{Hubert98} In our previous work on the influence of the surface anisotropy on the vortex core we compared in details different limiting regimes when magnetostatics can lead to effective anisotropy, see Ref.~\onlinecite{Pylypovskyi14} and references therein.

In order to validate our results for nanomagnets with the dipolar interaction, we also performed
few simulations with account of both surface anisotropy and dipolar interaction. In the current study we do not use \textsf{SLaSi} simulator for the dipolar interaction case, because of high numerical costs. Instead we make micromagnetic \textsf{OOMMF}\cite{oommf} simulations. As we already discussed above the modeling of the Bloch point is challenging due to singular magnetization distribution, leading typically to the mesh size dependence.\cite{Thiaville03} Nevertheless, according to the physical picture of the described effect, the key role in the \emph{dynamic} switching phenomenon plays the \emph{static} instability of the vortex core: there appears a spatial nonhomogeneity of the Bloch line, which causes breaks of Bloch line at the bottleneck places and finally results in the Bloch point mediated switching. That is why we perform \textsf{OOMMF} simulations of different static vortex distributions which precede switching. Numerically we model the Permalloy sample with account of dipolar interaction
without surface anisotropy (because the magnetostatics produces the effective ES one\cite{Caputo07b})
and additional EN surface anisotropy.\footnote{We simulate a nanodisks with cubic mesh of 2~nm. Magnetic fiels are directed along the disk sample. Materials parameters of Permalloy are used: exchange constant $A = 1.3\cdot 10^{-6}$~erg/cm, saturation magnetization $M_s = 68.7$~Oe (860~kA/m). The surface anisotropy is simulated as nonzero easy-axis anisotropy of value $|\beta_{\text{bulk}}|= 2.4\pi M_s^2$ in the cells on the face surfaces which corresponds to the surface anisotropy $|\beta_{\text{surf}}| = a|\beta_{\text{bulk}}| \approx 19.7$~Merg/cm$^2$. }

Similar to the Heisenberg magnet with EN surface anisotropy the vortex core in the nanodisk with magnetostatics and strong enough EN surface anisotropy has a pillow-shaped profile without external field, see Fig.~\ref{fig:oommf}a [data for $H=0$ and schematic (c)]. Under the action of the field there appears the curve bend of the vortex core profile. Under the action of strong enough field the vortex profile takes up a deformed pillow shape and barrel shape in the stronger field just before switching field, see
Fig.~\ref{fig:oommf}a [data for $H=-3$~kG (0.3~T) with schematic (d) and data for $H=-4.4$~kG (0.44~T)]. Even in the absence of additional ES surface anisotropy in the sample with large aspect ratio $ L/(2R) \gtrsim 1 $ the initial barrel-shaped vortex core deforms in the central part in the same case as it is shown in Fig.~\ref{fig:w-via-z-field}e. Such a bending is absent in the thin samples where the creation of volume magnetostatic charges significantly influnces to the surface ones. In the both cases of surface anisotropies switching starts from the places, where vortex core becomes more narrow under the action of external field: from the surfaces for EN and from the center for effective ES surface anisotropy.
Thus, the influence of the external field on the vortex profile and switching mechanism for the EN and ES surface anisotropies is in a good qualitative agrement with results of \textsf{SLaSi} simulations, see Fig.~\ref{fig:w-via-z-field} and wired core model, see Fig.~\ref{fig:wcm-field}.

\begin{acknowledgments}

All simulations results presented in the work were obtained using the computing clusters of Taras Shevchenko National University of Kyiv \cite{unicc} and Bayreuth University \cite{btrzx}.

\end{acknowledgments}

\appendix

\section{Equilibrium magnetization in film with surface anisotropy}
\label{app:m0}

Let us consider a film of thickness $L$ with constant of the easy-plane anisotropy $K > 0$ and constant of the surface anisotropy $K_s = \varkappa K$. The equilibrium magnetization distribution is determined by the Eqns.~\eqref{eq:BVP-tot}. In order to derive the magnetization distribution one can assume that $\vec m = \sqrt{1-m_0^2} \hat{\vec x} + m_0 \hat{\vec z}$ without lost of generality, where $m_0 = m_0(z,h)$. In this case Eqs.~\eqref{eq:BVP-tot} are reduced to
\begin{align} \label{eq:BVP-tot-3}
\ell^2 \dfrac{\partial^2 m_0}{\partial z^2} + h - m_0 = \dfrac{\ell^2 m_0}{\sqrt{1-m_0^2}} \dfrac{\partial^2}{\partial z^2}\sqrt{1-m_0^2},\\
\left[\ell\dfrac{\partial m_0}{\partial z} \pm \dfrac{\varkappa }{\sqrt{\lambda }} m_0 (1 - m_0^2)\right] \Biggr|_{z = \pm L/2} = 0.
\end{align}
The analytical solution can be found in assumption that $|\varkappa /\sqrt{\lambda }|\ll 1$. Assuming $m_0(z, h) = h + \frac{\varkappa }{\sqrt{\lambda }} y(z)$ the solution takes the form~\eqref{eq:m0-z}.

\section{Simulations with exchange anisotropy}
\label{sec:Heisenberg-SAex-numerics}

The surface anisotropy naturally appears in the Heisenberg ferromagnet with exchange anisotropy. \cite{Kosevich90} That is why we additionally simulate the system, described by the LLG equations~\eqref{eq:LLG} and Hamiltonian
\begin{equation*}\label{eq:exch-an-ham}
\begin{split}
 \mathscr H = &- J\mathcal{S}^2 \sum_{\vec{n},\vec{\delta }} \vec{m_{\vec{n}}} \cdot
\vec{m_{\vec{n}+\vec{\delta }}} + \frac{J_z\mathcal{S}^2}{2} \sum_{\vec{n},\vec{\delta }}
m_{\vec{n}} m_{\vec{n}+\vec{\delta }} \\
 & -  2\mu_BH\mathcal{S}\sum_{\vec{n}} m_{\vec n},
\end{split}
\end{equation*}
where $J_z > 0$ is a coefficient of the exchange anisotropy for easy-plane magnet. The sum in the second term in the Eq.~\eqref{eq:exch-an-ham} runs over 6 neighbors for volume sites and over 3--5 for surface sites which can be interpreted as a exchange-induced surface anisotropy due symmetry breaking of the lattice on the surface.

We use the same procedure of simulations as in previous sections with $J_z/J=0.005$, which gives the vortex polarity reversal under the action of the reduced magnetic field $h =2\mu_BH/(J_z\mathcal{S})=2.2$. The observed magnetization dynamics is qualitatively similar to the case of easy-plane volume anisotropy with EN surface anisotropy shown in Fig.~\ref{fig:EN-data}. In comparison with simulations with single-ion anisotropy the difference in vortex core magnetization is much smaller in this case. We compute the difference of dynamical polarities in the center and face surfaces of the sample $\Delta \mu^\text{EN}=|\mu_\text{top}-\mu_\text{center}| = 1.65$ (see Fig.~\ref{fig:EN-data}\,d) and $\Delta \mu^\text{Ex} = 0.032$ which is about 50 times weaker than EN surface anisotropy.

\section{Ring under the action of DC magnetic field}
\label{app:ring}

Let us consider a magnetic ring of inner radius $r_\text{in} \ge 0.3 \ell$ in the vortex state described by the Hamiltonian~\eqref{eq:H-total}. For such an inner radius without external fields the out-of-plane magnetization component vanishes and planar vortex appears.\cite{Kravchuk07} The Eq.~\eqref{eq:BVP-tot-1} takes the following form after substitution of the vortex anzats~\eqref{eq:vortex-phi}:
\begin{equation} \label{eq:initRing}
\ell^2 \nabla^2 m_z + h - m_z = \ell^2 \dfrac{m_z^2}{m_y}\nabla^2 m_y,
\end{equation}
where $m_y = \sqrt{1-m_z^2}\cos\phi$. Let us consider the case of weak fields $|h| \ll 1 $. We find the solution of Eq.~\eqref{eq:initRing} in the form
\begin{equation*}
 m_z = h m_r.
\end{equation*}
The Eq.~\eqref{eq:initRing} takes the form
\begin{subequations}\label{eq:BVPring}
\begin{equation}
 \ell^2 \left(\dfrac{\partial^2 m_r}{\partial z^2} + \dfrac{\partial^2 m_r}{\partial r^2} + \dfrac{1}{r}\dfrac{\partial m_r}{\partial r}\right) + m_r \left( \dfrac{\ell^2}{r^2} - 1 \right) + 1 = 0
\end{equation}
\begin{equation}\label{eq:bz}
\left(\dfrac{\partial m_r}{\partial z} \pm \dfrac{\varkappa a}{\ell^2} m_r \right) \Biggr|_{z = \pm L/2} = 0,
\end{equation}
\begin{equation}\label{eq:br}
 \dfrac{\partial m_r}{\partial r}\Biggr|_{r=r_\text{in},R} = 0.
\end{equation}
\end{subequations}For $\varkappa \neq 0$ the solution of~\eqref{eq:BVPring} can be found in the form
\begin{equation}\label{eq:mrring}
m_r = \sum_n R_n(r)Z_n(z)
\end{equation}

\begin{figure}
\begin{center}
\begin{tikzpicture}
\node at (0,0) {\includegraphics[width=\columnwidth]{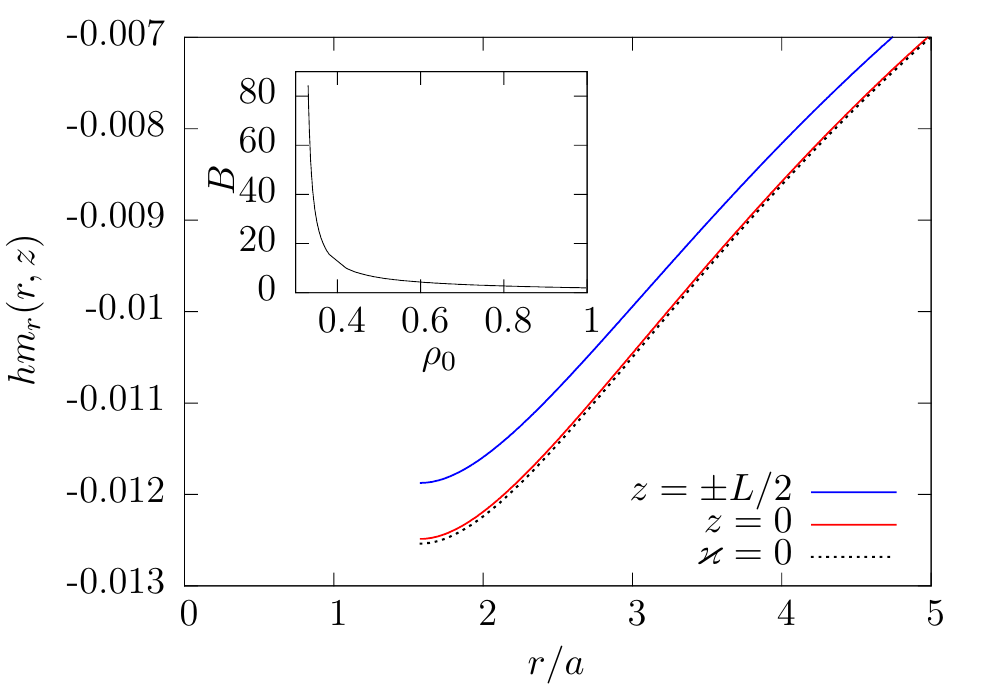}};

\end{tikzpicture}
\end{center}
\caption{(Color online) Magnetization distribution $ m_z = h m_r(r, z) $ in the center and on the face surfaces of the ring with $r_\text{in} = 1.58a$ and ES surface anisotropy under the action of the field $h = - 0.0009$ (4 harmonics of the Eq.~\eqref{eq:bvpringkappa} are taken into account). Dotted curve corresponds to the function $h m_{rh}$ (absence of surface anisotropy). Inset shows value of coefficient $B$ as function of $\rho_0 = \sqrt{1+\sigma_n^2} r_\text{in}/\ell$ %
$\varkappa = 0.1$, other parameters are the same as in Fig.~\ref{fig:wcm-field}.
}
\label{fig:mrring}
\end{figure}

which results in the following boundary-value problem:
\begin{subequations}\label{eq:bvpringkappa}
\begin{equation}
 \ell^2\dfrac{\mathrm d^2 Z_n}{\mathrm dz^2} + \sigma_n^2 Z_n = 0,
\end{equation}
\begin{equation}  \label{eq:mrkappa}
 \ell^2 \left(\dfrac{\mathrm d^2 R_n}{\mathrm dr^2} + \dfrac{1}{r}\dfrac{\mathrm d R_n}{\mathrm d r}\right) + R_n \left( \dfrac{\ell^2}{r^2} - 1 - \sigma_n^2 \right) + F_n = 0,
\end{equation}
\begin{equation}
\left(\ell \dfrac{\mathrm d Z_n}{\mathrm dz} \pm \dfrac{\varkappa}{\sqrt{\lambda}} Z_n\right) \Biggr|_{z = \pm L/2} = 0,
\end{equation}
\begin{equation}\label{eq:radial}
 \dfrac{\mathrm d R_n}{\mathrm d r}\Biggr|_{r=r_\text{in},R} = 0.
\end{equation}
\end{subequations}
where $F_n$ are defined as coefficients of expansion
\begin{equation} \label{eq:1expansion}
1 = \sum_n F_n Z_n(z).
\end{equation}
Eigenfunctions $Z_n$ have the form
\begin{equation*}
 Z_n(z) = \sigma_n \cos \dfrac{\sigma_n (z + L/2)}{\ell} + \dfrac{\varkappa}{\sqrt{\lambda}} \sin \dfrac{\sigma_n (z + L/2)}{\ell},
\end{equation*}
with dispersion equation
\begin{equation*}
 \tan 2\dfrac{\sigma_n L}{2\ell} = 2\sqrt{\lambda} \dfrac{\sigma_n \varkappa}{\lambda \sigma_n^2 - \varkappa^2},\quad n \in \mathbb{Z}_+.
\end{equation*}
For the case $\varkappa < 0$, the first eigenvalue $\sigma_0$ should be found as solution of
\begin{equation*}
\tanh \dfrac{\sigma_0 L}{2\ell} = - \dfrac{\varkappa}{\sqrt{\lambda} \sigma_0}
\end{equation*}
with eigenfunction
\begin{equation*}
 Z_0(z) = \sigma_0 \cosh \dfrac{\sigma_0 (z + L/2)}{\ell} + \dfrac{\varkappa}{\sqrt{\lambda}} \sinh \dfrac{\sigma_0 (z + L/2)}{\ell}.
\end{equation*}
Solution of the Eq.~\eqref{eq:mrkappa} can be written in the following form
\begin{equation} \label{eq:radialHarm}
\begin{split}
 R_n(\rho) = & \dfrac{F_n}{1+\sigma_n^2} \left[ \mathrm K_i(\rho)\int_{1}^\rho \mathrm L_i(\rho)\rho \mathrm d\rho + B\mathrm K_i(\rho) \right.\\
 &\left.+ \mathrm L_i(x) \int_{\rho}^\infty \mathrm K_i(\rho)\rho \mathrm d\rho\right],
 \end{split}
\end{equation}
where $\rho = \sqrt{1+\sigma_n^2} r/\ell$, $\mathrm L_i(\bullet)$ and $\mathrm K_i(\bullet)$ are real valued solutions of modified Bessel equation of purely imarginal order with Wronskian $\mathcal W\{\mathrm K_i(\zeta), \mathrm L(\zeta)\} = 1/\zeta$.\cite{Dunster90,Gil04} Constant $B$ can be found from the boundary conditions~\eqref{eq:radial} at $r = r_\text{in}$:
\begin{equation}
\begin{split}
B(\rho _0) & = - \left[ \int_{1}^\rho \mathrm L_i(\rho)\rho \mathrm d\rho + \frac{\mathrm d \mathrm L_i/\mathrm d\rho}{\mathrm d \mathrm K_i/\mathrm d\rho} \int_{\rho}^\infty \mathrm K_i(\rho)\rho \mathrm d\rho \right]\Biggr|_{\rho=\rho_0}\\
& \approx B^{\text{fit}}(\rho _0)=\dfrac{\rho_0 + 0.557}{0.99 \rho_0 - 0.316},
\end{split}
\end{equation}
where $\rho_0 = \sqrt{1+\sigma_n^2}r_\text{in}/\ell$. The interpolation function $B^{\text{fit}}(\rho _0)$ fits numerically calculated dependence $B(\rho _0)$ in the range $\rho _0\in [0.34,0.8]$ with an accuracy of about $5\%$. Functions $m_r(r)$ and $B(\rho_0)$ are shown in Fig.~\eqref{fig:mrring}. In the absence of the surface anisotropy the solution of Eqns.~\eqref{eq:BVPring} can be found in the form $m_r = m_{rh}(r)$ which is the solution of the following equation:
\begin{equation} \label{eq:mrhom}
 \ell^2 \left(\dfrac{\mathrm d^2 m_{rh}}{\mathrm dr^2} + \dfrac{1}{r}\dfrac{\mathrm d m_{rh}}{\mathrm d r}\right) + m_{rh} \left( \dfrac{\ell^2}{r^2} - 1 \right) + 1 = 0
\end{equation}
with boundary condition~\eqref{eq:br}. Its solution can be obtained from Eq.~\eqref{eq:radialHarm} by replacing $R_n$ to $m_{rh}$ and substitution $F_n = 1$ and $\sigma_n = 0$.

For the calculations in scope of the reduced wired core model we use 3 harmonics for smallest $L$ and up to 7 for the most thick samples.

\section{Reduced Wired Core Model}
\label{app:core}

\begin{figure}
\begin{center}
\includegraphics[width=\columnwidth]{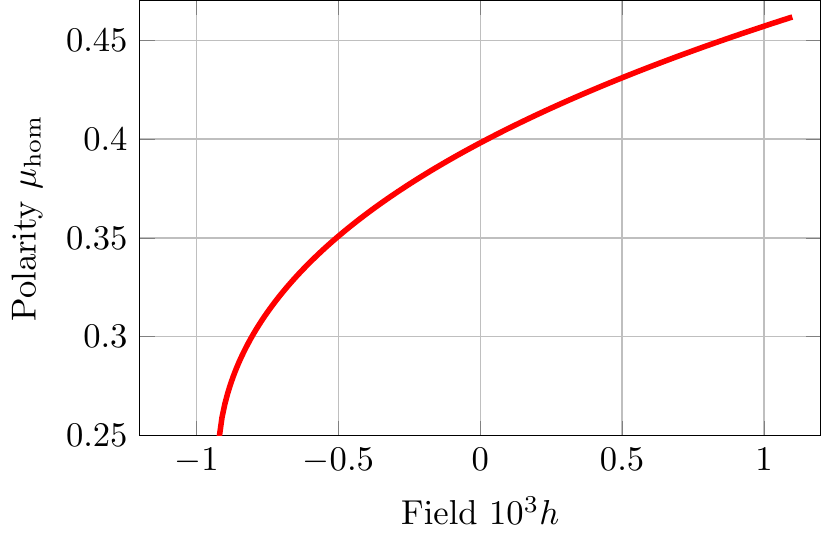}
\end{center}
\caption{(Color online) Equilibrium polarity $\mu_\text{hom}(h)$ for the discrete homogeneous model~\eqref{eq:Hwcm1} ($\varkappa_n=0$, $n=\overline{1,N}$) for $\lambda=20$.
}
\label{fig:polarityInField}
\end{figure}

The Hamiltonian \eqref{eq:H-total} with account of the anisotropy term \eqref{eq:H-EP+SA-2} can be rewritten in the following form:
\begin{equation*}
\begin{split}
\mathcal H_c =&  - J\mathcal{S}^2 \sum_{\vec n, \vec \delta } \Bigl[m_{\vec n}m_{\vec n+\vec \delta } \\
 &+ \sqrt{(1-m_{\vec n}^2)(1-m_{\vec n+\vec\delta }^2)}\cos(\phi_{\vec n}-\phi_{\vec n+\vec\delta }) \Bigr] \\
 & + \sum_{\vec n=(n_x,n_y,n)} \left( \frac{K_n\mathcal{S}^2}{2} m_{\vec n}^2 - 2\mu_B H\mathcal{S} m_{\vec n} \right).
\end{split}
\end{equation*}
By incorporating the reduced vortex core ansatz~\eqref{eq:freefixedSpins} one get
\begin{equation*}
\begin{split}
\mathcal H_c = & \sum_{n=1}^N 4K\mathcal{S}^2 \left[ - \frac{J}{K}\mu_n^2 - \frac{2J}{K}m_b\mu_n +  \varkappa_n\frac{\mu_n^2}{2} \right. \\
 & \left.- \frac{4J}{\sqrt{5}K} \sqrt{(1-m_b^2)(1-\mu_n^2)}\cos\psi_n + \frac{\mu_n^2}{2} - \mu_nh \right] \\
 & - 4J\mathcal{S}^2 \sum_{n=1}^{N-1} \left[ \mu_n\mu_{n+1} + \sqrt{(1-\mu_{n}^2)(1-\mu_{n+1}^2)} \right. \\
 & \times \cos(\psi_n - \psi_{n+1}) \biggr].
\end{split}
\end{equation*}
Taking into account that $\varkappa_n = 0$ for $n = \overline{2,N-1}$, $\varkappa_1=\varkappa_N=\varkappa$, $\lambda = \ell^2/a^2$, $\lambda_n = J/K_n \equiv \lambda/(1+\varkappa_n)$ one obtains the Eq.~\eqref{eq:Hwcm1}. The temporal evolution of $\mu_n(\tau )$ and $\psi_n(\tau )$ is governed by the Eq~\eqref{eq:LLG}, which results in
\begin{equation} \label{eq:llg-wcm}
\begin{split}
\frac{\mathrm{d}\mu_n}{\mathrm{d}\tau } & = \dfrac{1}{1+\varepsilon^2} \frac{\partial \mathscr H_c}{\partial \psi_n} - (1-\mu_n^2)\dfrac{\varepsilon }{1+\varepsilon^2} \frac{\partial \mathscr H_c}{\partial \mu_n},\\
\frac{\mathrm{d}\psi_n}{\mathrm{d}\tau } & = - \dfrac{1}{1+\varepsilon^2}\frac{\partial \mathscr H_c}{\partial \mu_n} - \frac{1}{1-\mu_n^2}\dfrac{\varepsilon }{1+\varepsilon^2} \frac{\partial \mathscr H_c}{\partial \psi_n}.
\end{split}
\end{equation}
The problem of magnetization dynamics in discrete model can be analysed numerically. For analytical approach we replace the discrete functions $\mu_n$ and $\psi_n$ using their continuous analogues $\mu (z)$ and $\psi (z)$ and replace summation by integration. The volume contribution has the form
\begin{subequations} \label{eq:wcm-energy}
\begin{equation}
\begin{split}
\mathscr E_v[\mu] = & \frac{4}{a}\int_{-L/2}^{L/2} \left[ (1-2\lambda )\frac{\mu^2}{2} - (h + 2\lambda m_b ) \mu \right. \\
 & \left.- \frac{4\lambda }{\sqrt{5}}\sqrt{(1-m_b^2)(1-\mu^2)}\cos\psi \right]\mathrm dz \\
 & + 2\lambda a \int_{-L/2}^{L/2} \left[ \frac{(\mu')^2}{1-\mu^2} + (1-\mu^2)(\psi')^2 \right] \mathrm dz,
\end{split}
\end{equation}
with $m_0 \equiv m_0(z, h)$ defined in Eq.~\eqref{eq:m0-z}. The surface contribution is
\begin{equation}
\mathscr E_s^\pm[\mu] = 2 \varkappa \mu^2 \Biggr|_{z=\pm L/2}.
\end{equation}
\end{subequations}
We assume that the deviation from the equilibrium dynamical polarity in absence of the surface anisotropy is small and $\mu (z) = \mu_0(h) + \frac{\varkappa }{\sqrt{\lambda }} x(z)$ with $|\varkappa/\sqrt{\lambda }| \ll 1$, where $\mu_\text{hom}(h)$ can be found numerically as solution of the following equation
\begin{equation} \label{eq:muhomh}
\begin{split}
\dfrac{1}{2} (1-2\lambda )\mu_\text{hom}^2 & - h(1+2\lambda m_{rh})\mu_\text{hom} \\
 & = \dfrac{4\lambda }{\sqrt{5}}\sqrt{(1-h^2m_{rh}^2)(1-\mu_\text{hom}^2)},
\end{split}
\end{equation}
see Fig.~\ref{fig:polarityInField}. Taking $\psi (z) = 0$ and varying the functional~\eqref{eq:wcm-energy} one can find the equilibrium distribution of $\mu$ as solution of the boundary-value problem~\eqref{eq:wcm-bvp}
where
\begin{subequations} \label{eq:wcm-coefficients}
\begin{equation}
	A_1 = \left(2 - \dfrac{1}{\lambda } \right)(1 - \mu_\text{hom}^2) - \dfrac{4}{\sqrt{5(1 - \mu_\text{hom}^2)}},
\end{equation}
\begin{equation}
\begin{split}
A_2(z) = & \dfrac{1 - \mu_\text{hom}^2}{5\varkappa \sqrt{\lambda }}\Biggl[ 5h - 10\lambda m_b(z) \\
 & + \mu_\text{hom} \left( 5 - 10\lambda + 4\lambda \sqrt{\dfrac{5}{1 - \mu_\text{hom}^2}} \right) \Biggr].
\end{split}
\end{equation}
\end{subequations}

%
%
%

\end{document}